\documentclass[fleqn]{2020SCGE}
\usepackage{graphicx}
\usepackage{hyperref}

\newcommand{\mpch}{\>h^{-1}{\rm {Mpc}}}

\newcommand{\kmsmpc}{\>{\rm km}\,{\rm s}^{-1}\,{\rm Mpc}^{-1}}

\usepackage{color}
\newcommand{\adb}[1]{\textcolor{blue}{ #1}} 

\UseRawInputEncoding

\begin{document}

\ensubject{subject}

\ArticleType{Article}
\SpecialTopic{SPECIAL TOPIC: }
\Year{2020}
\Month{xxx}
\Vol{xx}
\No{x}
\DOI{xx}
\ArtNo{000000}
\ReceiveDate{xxx xxx, 2020}
\AcceptDate{xxx xxx, 2020}

\title{The clustering of galaxies in the DESI imaging legacy surveys DR8: 
I. the luminosity and color dependent intrinsic clustering}{}

\author[1]{Zhaoyu Wang}{}
\author[1]{Haojie Xu}{}
\author[1,2]{Xiaohu Yang}{{xyang@sjtu.edu.cn}}
\author[1,2]{Yipeng Jing}{{ypjing@sjtu.edu.cn}}
\author[3,4]{Kai Wang}{}
\author[5]{Hong Guo}{}
\author[6]{\\Fuyu Dong}{}
\author[1]{Min He}{}

\AuthorMark{Wang et al.}

\AuthorCitation{Wang, Z., Xu, H., Yang, X. et al.}

\address[1]{Department
of Astronomy, School of Physics and Astronomy, and Shanghai Key Laboratory for Particle Physics and Cosmology, \\
Shanghai Jiao Tong University, Shanghai {\rm 200240}, China}
\address[2]{Tsung-Dao Lee Institute and Key Laboratory for
    Particle Physics, Astrophysics and Cosmology, Ministry of Education, \\
    Shanghai Jiao Tong University, Shanghai {\rm 200240}, China}
\address[3]{Department of Astronomy, Tsinghua University, Beijing {\rm 100084}, China}
\address[4]{Department of Astronomy, University of Massachusetts Amherst, {\rm MA 01003}, USA}
\address[5]{Key Laboratory for Research in Galaxies and Cosmology, Shanghai Astronomical Observatory, Shanghai {\rm 200030}, China}
\address[6]{School of Physics, Korea Institute for Advanced Study (KIAS), 85 Hoegiro, Dongdaemun-gu, Seoul, {\rm 02455}, Republic of Korea
}

\abstract{In a recent study, we developed a method to model the impact of photometric
	redshift uncertainty on the two-point correlation function (2PCF). In this method, we can
	obtain both the intrinsic clustering strength and the photometric redshift errors simultaneously 
	by fitting the projected 2PCF with two integration depths along the line-of-sight. 
	Here we apply this method to the DESI Legacy Imaging Surveys Data Release 8 (LS DR8), the largest galaxy sample currently available. 
	We separate galaxies into 20 samples in 8 redshift bins from $z=0.1$ to
	$z=1.0$, and a few $\rm z$-band absolute magnitude bins, with $M_{\rm z} \le -20$. These galaxies are further separated into red and blue sub-samples according to their $M^{0.5}_{\rm r}-M^{0.5}_{\rm z}$ colors.  We measure the projected 2PCFs for all these galaxy (sub-)samples, and fit them using our
	photometric redshift 2PCF model. We find that the photometric redshift errors are smaller in 
	red sub-samples than the overall population. On the other hand, there might be some systematic photometric redshift errors in the blue sub-samples, so that some of the sub-samples show significantly enhanced 2PCF at large scales.  Therefore, focusing only on the red and all (sub-)samples, we find that the biases of galaxies in these (sub-)samples show clear color, redshift and luminosity dependencies, in that red  brighter galaxies at higher redshift are more biased than their bluer and low redshift counterparts. Apart from the best fit set of parameters, $\sigma_{z}$ and $b$, from this state-of-the-art photometric redshift survey, we obtain high precision intrinsic clustering measurements for these 40 red and all galaxy (sub-)samples. These measurements on large and small scales hold important information regarding the cosmology and galaxy formation, which will be used in our subsequent probes in this series. 
	}



\maketitle


\begin{multicols}{2}
\section{Introduction}\label{sec:intro}

Past decades have seen the flourishing of large galaxy redshift surveys. A large number of surveys, such as the Las Campanas Redshift Survey (LCRS)
\cite{LCRS}, 
the 2 degree Field Galaxy Redshift Survey (2dFGRS)
\cite{2dFGRS}, \Authorfootnote
the Sloan Digital Sky Survey (SDSS) \cite{SDSS},
the DEEP2 Galaxy Redshift Survey (DEEP2) \cite{deep2_clustering1,
deep2_clustering2}, and the VIMOS Public Extragalactic
Redshift Survey (VIPERS) \cite{vipers_clustering} have provided tremendous
measurements of the galaxy spatial distributions, which enable us to explore precise cosmological parameters. The Baryon Acoustic Oscillations (BAO) \cite{2010dken.book..246B, 2012MNRAS.427.2132P, 
2014MNRAS.441...24A, 2015PhRvD..92l3516A, 2019MNRAS.484.3818S},  gravitional lensing \cite{2001PhR...340..291B, 2008ARNPS..58...99H, 2014MNRAS.442.2017V, 2015MNRAS.446.1356H, 2016MNRAS.456.2301W,  2015JCAP...01..024Z,2017ApJ...836...38L, 2018ApJ...862....4L, 2019ApJ...874....7D}, power spectrum
\cite{1994ApJ...426...23F, 2002MNRAS.330..506H,  2002MNRAS.335..887T,
    2004ApJ...606..702T, 2011MNRAS.416.3017B, 2016ApJ...833..287L,
    2001MNRAS.327.1297P, 2007MNRAS.381.1053P, 2010MNRAS.401.2148P} and correaltion functions (such as two-point correlation
functions (2PCF) and three-point correlation functions (3PCF)
\cite{2005MNRAS.361..824G, 2009ApJ...698..479G, 2009ApJ...702..425G}  are
often used to depict the universe.

The 2PCF, a second-order moment statistics, or its integrated version, \adb{the} projected
2PCF, are usually employed to characterize the clustering strength of galaxies.
Numerous studies with galaxy
samples at low-$z$ and intermediate-$z$ reveal strong correlations between
clustering strength and galaxy properties, such as color, luminosity, spectral type \cite{jing98, 2004MNRAS.350.1153Y,
    2005ApJ...633..560E, 2005ApJ...630....1Z, licheng06, zehavi11, guo14,
guo15, 2016ApJ...833..241S, 2018ApJ...858...30G, Xu18} and so on.

To explain the correlation between clustering strength and galaxy properties,
some statistical model based on halo-galaxy connection are proposed, such as
the halo occupation distribution (HOD) \cite{jing98,
    2002ApJ...575..587B,zheng05, zehavi11, guo16, 2018MNRAS.478.2019Y,
    2018ApJ...858...30G, 2015MNRAS.454.1161Z, 2016MNRAS.457.4360Z,
2018MNRAS.476.1637Z}, the conditional luminosity function model (CLF) 
\cite{yang03, 2006MNRAS.365..842C, 2007MNRAS.376..841V, yang12,
    2015ApJ...799..130R} and halo abundance matching \cite{2006ApJ...647..201C,
guo16}. 
These models transform galaxy clustering measurements to the 
informative, physical relation between galaxies and dark matter halos, which encodes
the complex physics of galaxy formation and evolution, and are very helpful in diagnosing galaxy 
formation theories. 
Under the HOD/CLF framework, tight constraints on the cosmology
can also be achieved by a combination of galaxy clustering,
dynamical clustering measurements (i.e., pairwise radial velocity
dispersion of galaxies), and galaxy-galaxy lensing \cite{2007ApJ...667..760Z,2009MNRAS.394..929C,2013MNRAS.430..767C}.

Most of the previous work on galaxy clustering focus on galaxies with
spectroscopic redshift (spec$z$), which is relatively easy to measure in local
universe. For high-$z$ galaxies, due to the observational limitation and
measurement difficulty, one can only get a relatively small galaxy sample in
a small volume, such as DEEP2 \cite{deep2_clustering1, deep2_clustering2}, 
zCOSMOS \cite{2007ApJS..172...70L}. Hence, the
2PCF measurement will become noisy due to the Poisson noise and cosmic
variance, especially on large scales. 

In general, there are two types of approaches to infer galaxy-halo connection at these high redshifts to increase the signal-to-noise ratios of the clustering measurements. One is to measure the cross-correlation between the photometric and spectroscopic samples \cite{Masjedi06,2009MNRAS.399.2279M,2011ApJ...731..117H,wangwenting11} and the other is to directly model the angular galaxy clustering measurements \cite{2012A&A...542A...5C,HSC_clustering16, 2018PASJ...70S..11H, 2018ApJ...853...69C, 2018PASJ...70S..33H}.
However, both approaches have their own limitations. The cross-correlation method is limited by the size of the spectroscopic sample and needs careful treatment of the interlopers, while the galaxy-halo connection in the angular clustering method is less well constrained due to the lack of redshift information and hard to probe the redshift evolution of galaxies.

To overcome these problems, we proposed a new method in \cite{2019ApJ...879...71W} (hereafter W19) 
to directly measure the projected 2PCFs from galaxies
with only photometric redshifts. In W19, we constructed a realistic mock
light-cone from N-body simulation where the galaxies are populated using HOD
model. The tests showed that our method can not only recover the 2PCF, but also
constrain the photometric redshift uncertainties, which is not achievable for
previous methods \cite{Eisenstein03, wangwenting11, HSC_clustering16}.

In this paper, we will apply the method developed in W19 to the DESI Legacy Imaging Surveys.  
Here we aim to obtain the intrinsic (i.e. corrected for the impact of photometric redshift errors) projected 2PCFs  for sets of galaxy (sub-)samples in different luminosity and
redshift bins. In this regards, before the availability of the spectroscopic redshifts from the subsequent DESI observations, we can already have clustering measurements that are much accurate than before for the current galaxy formation and cosmological studies. The structure of this paper is organized as follows. First, we introduce the DESI Legacy Imaging Surveys data and our clustering measurement method in
\S\ref{sec:data}. Then photometric redshift modeling of the 2PCF is presented in \S\ref{sec:method}. We
show the main results in \S\ref{sec:results} and the summary in
\S\ref{sec:summary}. Throughout this paper, we assume a $\Lambda$CDM cosmology
that are consistent with the Planck 2018 results \cite{Planck2018}: $\Omega_{\rm m} = 0.315$, $\Omega_{\Lambda} = 0.685$, $n_{\rm s}=0.965$, $h=H_0/(100 \kmsmpc) = 0.674$ and $\sigma_8 = 0.811$.

\section{Galaxy Samples and Clustering Measurements}\label{sec:data}

We investigate the luminosity and color dependence of intrinsic
galaxy clustering with the DESI Legacy Imaging Surveys Date Release 8
(LS DR8; \cite{DESI_image}). LS DR8 provides target catalogs for The
Dark Energy Spectroscopic Instrument (DESI; \cite{DESI}), the
next generation spectroscopic survey for measuring the dark energy
effect on the expansion of the universe. All the galaxy positions
and properties are extracted from here
\footnote{\url{https://www.legacysurvey.org/dr8/files/}}.

Following the target selection criteria lay out in Yang et al. (2020) \cite{2020arXiv201214998Y}, we
select out galaxies by morphological classification
as type {\it REX}, {\it EXP}, {\it DEV}, and {\it COMP} 
from TRACTOR \cite{2016ascl.soft04008L} fitting results \footnote{In the Tractor fitting procedure, {\it REX} stands for round exponential galaxies with a variable radius,  {\it DEV} for deVaucouleurs profiles (elliptical galaxies), {\it EXP} for exponential profiles (spiral galaxies), and {\it COMP} for composite profiles that are deVaucouleurs plus exponential (with the same source center).}.
We also remove those objects at low galactic latitudes and in the vicinity of masked pixels and brights stars (see the selection 
details in section 2.1 \cite{2020arXiv201214998Y}).

The corresponding photometric redshift (photoz hereafter) 
catalog is adopted from the Photometric Redshifts for the Legacy Surveys (PRLS \footnote{https://www.legacysurvey.org/dr8/files/\#photometric-redshift-files-8-0-photo-z-sweep-brickmin-brickmax-pz-fits}\cite{2020arXiv200106018Z}), who employs
random forest algorithm \cite{Breiman2001} to estimate the photoz.
Same as the procedures done in \cite{2020arXiv201214998Y}, we select galaxies with ${\rm z\le 21}$ 
and  $ 0 < $ z\_phot\_median $\le 1$, where the z\_phot\_median is the median value of photoz probability distribution function. 
For those provided with spectroscopic redshift\footnote{The spectroscopic redshifts are from BOSS, SDSS, WiggleZ, GAMA, COSMOS2015, VIPERS, eBOSS, DEEP2, AGES, 2dFLenS,VVDS, and OzDES. See details in section 3.2 \cite{2020arXiv200106018Z} } in the catalog, 
we replace the photometric redshift with the spectroscopic redshift.
We include also additional spectroscopic redshifts matching from the 2MASS Redshift Survey (2MRS \cite{2012ApJS..199...26H}), 6dF Galaxy Survey Data Release 3 (6dFGRS \cite{2009MNRAS.399..683J}), and 2dF Galaxy Redshift Survey (2dFGRS \cite{2dFGRS}).

All the magnitudes and color used in the work are in the AB system and corrected for Galactic extinction. 
We convert apparent magnitude to absolute magnitude via the following equation,
\begin{equation} \label{eq:magnI}\nonumber
M_{\rm x} - 5\log h = 
m_{\rm x}  - {\rm DM}(z) - {\rm K^{0.5}_x}\,
\end{equation}
where ${\rm x} \equiv {r,z}$ and DM($z$) is the distance module corresponding to redshift $z$, 
\begin{equation} \label{eq:distmeas}\nonumber
{\rm DM}(z) = 5 \log D_L(z) + 25
\end{equation}
with ${\rm D_L}$($z$) being the luminosity distance in ${\rm \mpch}$.
${\rm K^{0.5}_x}$ represents the ${\rm K}$-correction in {x}-band to sample median redshift $z\sim0.5$.
Here ${\rm K}$-correction is obtained for each galaxy according to its three optical ({\it grz}) bands, two mid-infrared (W1 3.4$\mu$m, W2 4.6$\mu$m) bands photometries and the photoz information using the `Kcorrect' model (eg. v4\_3) described in \cite{2007AJ....133..734B}. Those who are interested can refer to \cite{2020arXiv201214998Y} to see an illustration of the $z$-band K-correction distributions as a function of redshift in their Fig. 2. In general, the photoz error can impact both the K-correction as well as the luminosity distance measurements, and hence the absolute magnitude estimation. However, since the overall quality of the photoz estimation in \cite{2020arXiv200106018Z} is remarkably 
good, the distance error in general is smaller than 5\%. We neglect its impact on the absolute magnitude estimation for each galaxy.

\subsection{Galaxy Sample Construction with Luminosity, Redshift, and Color}
\label{subsec:binning}

To investigate first the luminosity dependence of intrinsic galaxy clustering
and photoz errors, we construct galaxy samples in bins of redshift and 
galaxy luminosity, in light of
the fact that both galaxy bias and photoz quality are primarily dependent on galaxy 
luminosity and redshift. Zhou et al.\cite{2020arXiv200106018Z} suggested that galaxy photoz overall are
most accurate with ${\rm z}<$21 (see their Appendix B for details), we therefore adopt 
the ${\rm z}$ band absolute magnitude ($M^{0.5}_{\rm z}-5\log h$) as our galaxy luminosity indicator.
Since the photoz error is much larger than specz error, a relatively large 
redshift bin ($\Delta z = 0.2$ in this work) for 
galaxy samples is preferred to mitigate the effect due to galaxy interlopers from 
neighboring redshift bins. In principle, we divide the galaxies with 
absolute magnitude bin width $\Delta (M^{0.5}_{\rm z}-5\log h) = 1 $ mag, 
but the enormous amount of galaxies in some galaxy samples (more than 20 million galaxies) 
allows us to probe more subtle luminosity dependence. For this reason, in those samples we further divide galaxies
into bin with $\Delta (M^{0.5}_{\rm z}-5\log h) = 0.5 $ mag, regardless of galaxy number counts in these samples afterwards.

\begin{figure}[H]
\centering
\includegraphics[width =0.47\textwidth]{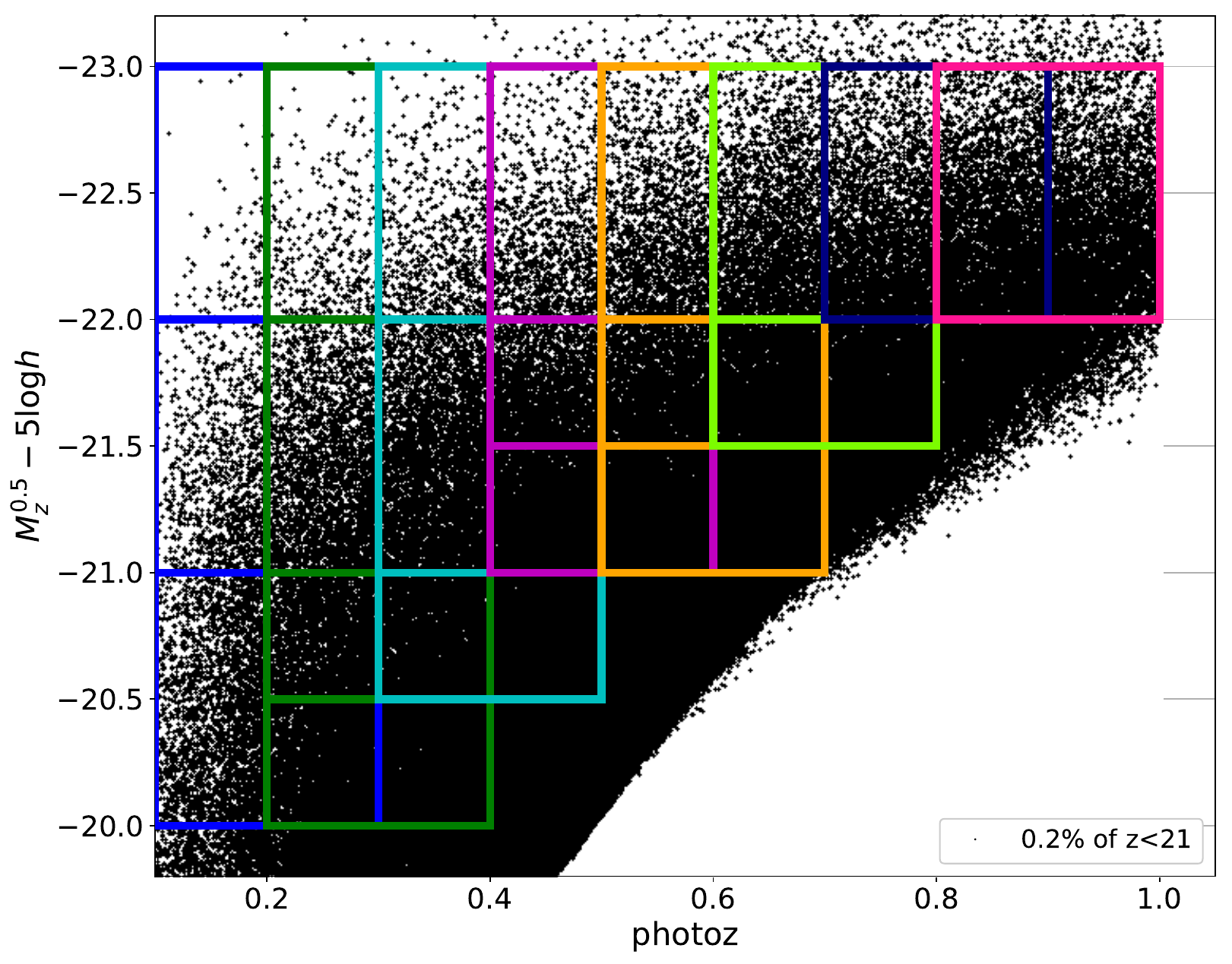}
\caption{Construction of galaxy samples in bins of $M_{\rm z}^{0.5}-5\log h$ and photoz.
Each patch represents a luminosity volume-limit sample, with different
color corresponding to different redshift bin. All luminosity samples
span redshift with $\Delta z = 0.2$ and the artificial z$\le$21 flux limit we
choose determines how faint we construct the samples.
As discussed in 
section ~\ref{subsec:binning}, some luminosity bins have  
absolute magnitude bin width $\Delta (M^{0.5}_{\rm z}-5\log h) = 0.5 $ mag, depending on
the galaxy numbers. The maximum luminosity goes to $M_{\rm z}^{0.5}-5\log h = -23$ for
all redshift bins. To avoid saturation, here we only plot the distribution of randomly selected
  0.2\% galaxies.}
\label{fig:sample}
\end{figure}

\begin{figure*}[!t]
\centering
\includegraphics[height=0.5\textwidth,width=0.95\textwidth]{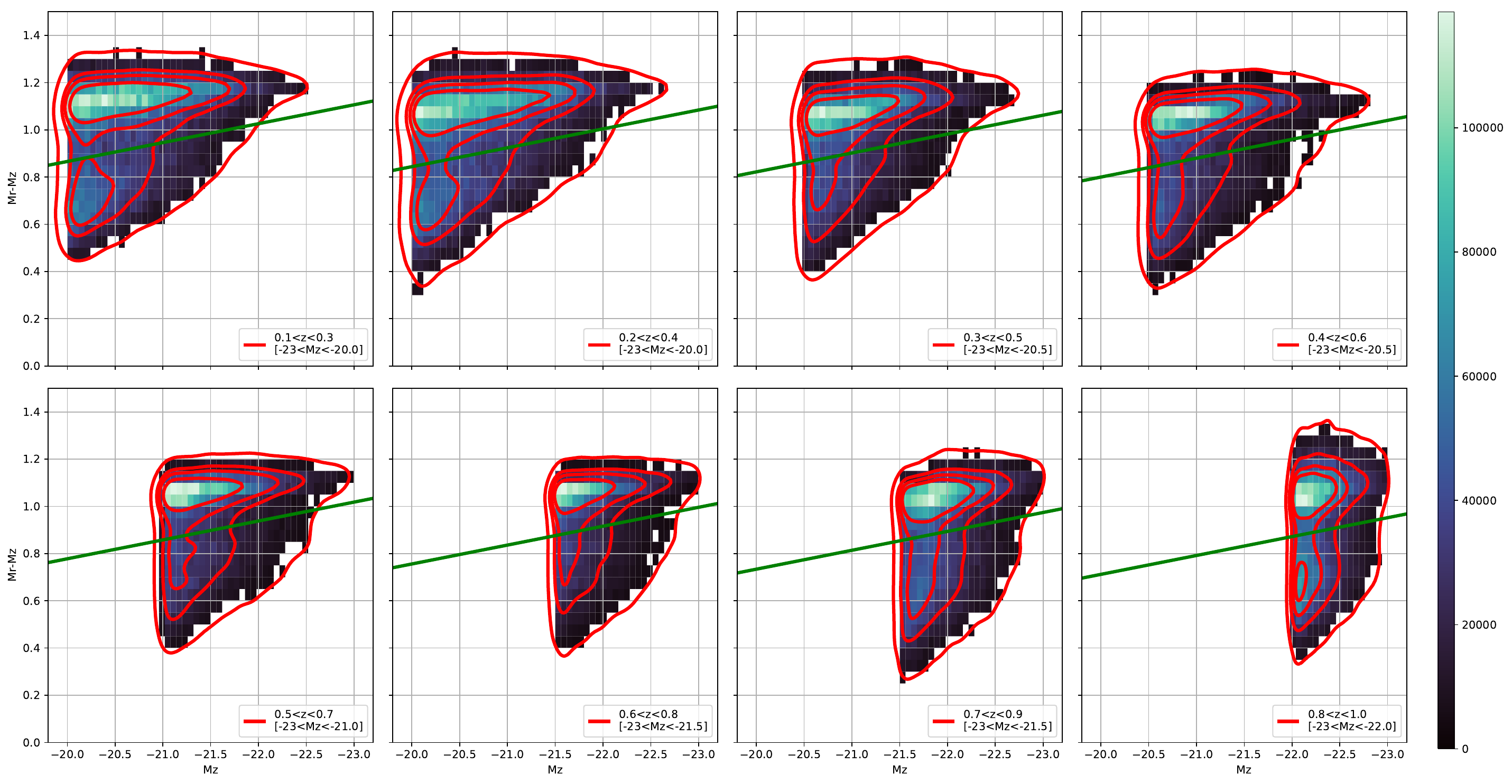}
\caption{K-corrected $M^{0.5}_{\rm r}-M^{0.5}_{\rm z}$ vs. $M^{0.5}_{\rm z}-5\log h$ for 
a random subsample (50,000 galaxies in each panel) of 
volume-limited luminosity samples as a function of redshift. 
The 2D histogram in each panel 
shows galaxy number counts
in fine bins of color and magnitude, where color codes
the galaxy number counts per magnitude per color, i.e.,
${\rm d^2Ng}$/d$(M^{0.5}_{\rm z}-5\log h)/{\rm d}(M^{0.5}_{\rm r}-M^{0.5}_{\rm z})$. The contours gives a less cluttered
view of the 2D histogram. The green solid line (Eq.\ref{eq:color}) divides the luminosity
samples into red/blue (or passive/star-forming) subsamples,
taking into account of the dependence on luminosity and redshift. 
Note that $M_{\rm z}$ shown in the legends refers to $M^{0.5}_{\rm z}-5\log h$.
}
\label{fig:CMD}
\end{figure*}

With the above consideration taken into account, the redshift of the galaxy samples 
ranges from centering at $z=0.2$ to at $z=0.9$, the galaxy samples are constructed in a volume-limited way.
We end up with 20 galaxy samples defined by bins in $M^{0.5}_{\rm z}-5\log h$ and photoz and covering the galaxy
luminosity-redshift diagram, which is shown in Fig.~\ref{fig:sample}. 
More detailed galaxy sample information 
is summarized in Table \ref{table:sample}.

In addition to luminosity and redshift, galaxy clustering depends also
on color, spectra type, morphology and surface brightness \cite{2002ApJ...571..172Z,2005ApJ...630....1Z, zehavi11, Xu18}. These properties are strongly correlated with
each other and display a similar change of clustering when dividing
galaxies based on them \cite{2002ApJ...571..172Z}. In this work, we choose color
since it is the immediately available quantity and the least vulnerable
one to measurement uncertainties. Besides this, Blanton \cite{2005ApJ...629..143B} found that 
luminosity and color are the two most predictive quantities for the
galaxy local density, with weak residual dependence on morphology
or surface brightness once the luminosity and color is fixed.

Following the color division practice at low redshift universe,
we plot the color-magnitude diagram constructed from a random
down-sampling (50,000) of previous volume-limited luminosity samples 
as a function of redshift, shown in Fig.~\ref{fig:CMD}. The contours
demonstrate two galaxy populations with a tight red sequence and a 
lose blue cloud in all redshift, which allows us to divide the galaxy
into red/blue (or passive/star-forming) subsamples by the following equation,

\begin{equation}\label{eq:color}
    M^{0.5}_{\rm r}-M^{0.5}_{\rm z} = -0.8 - 0.08*(M^{0.5}_{\rm z}-5\log h)-0.22*(z-0.5)
\end{equation}
where the $M^{0.5}_{\rm r}-5 \log h$ is the r-band absolute magnitude.
It seems that $M^{0.5}_{\rm r}-M^{0.5}_{\rm z}$ color serves well for the passive/star-forming subsamples division.
It might due to the fact that the 4000$\AA$ break moves to r-band at $z\sim0.5$, where the
K-correction is performed. Note that at relatively
high redshift, the dust could play a big role in linking the color and star-forming
activity such that star-forming galaxies may exhibit a red color. 
To what extend this effect could contaminate our color subsamples needs a further
study and it exceeds the scope of this work.

\begin{table*}[!t]
\centering
\footnotesize
\caption{The (sub-)samples selection criteria, number, number density, the related $\sigma_{z}$ and galaxy bias best fitting results for all and red galaxies. 
}\vspace*{1.5mm}
\label{table:sample}
\tabcolsep 6pt 
\begin{tabular*}{\textwidth}{lcccccc}
\toprule\vspace*{1.5mm}
Sample & redshift & absolute magnitude & numbers (all/red) & number density (all/red) & $\sigma_{z}$ (all/red) &  bias (all/red)\\
 & & $M_{\rm z}^{0.5}-5\log h$ &  & $10^{-4}(\mpch)^{-3}$ &
\\\hline\vspace*{1.5mm}
1& [0.1,0.3] & [-21.0,-20.0] & 6636956/4162406 & 64.6715/40.5591       & $0.0149^{+0.0005}_{-0.0004}$/$0.0134^{+0.0004}_{-0.0003}$ & $1.0591^{+0.0114}_{-0.0107}$/$1.2402^{+0.0103}_{-0.0099}$ \\\hline\vspace*{1.5mm}
2& [0.1,0.3] & [-22.0,-21.0] & 2502266/1908969  & 24.3824/18.6013       & $0.0110^{+0.0002}_{-0.0002}$/$0.0103^{+0.0002}_{-0.0002}$ & $1.2065^{+0.0066}_{-0.0064}$/$1.3202^{+0.0067}_{-0.0065}$ \\\hline\vspace*{1.5mm}
3& [0.1,0.3] & [-23.0,-22.0] & 264990/226065     & 2.5821/2.2028   & $0.0076^{+0.0002}_{-0.0002}$/$0.0069^{+0.0003}_{-0.0003}$ & $1.6373^{+0.0091}_{-0.0092}$/$1.7131^{+0.0096}_{-0.0097}$ \\\hline\vspace*{1.5mm}
4& [0.2,0.4] & [-20.5,-20.0] & 7648188/4408031 & 38.2338/22.036       & $0.0242^{+0.0020}_{-0.0016}$/$0.0201^{+0.0010}_{-0.0008}$ & $1.1045^{+0.0375}_{-0.0313}$/$1.2765^{+0.0249}_{-0.0217}$ \\\hline\vspace*{1.5mm}
5& [0.2,0.4] & [-21.0,-20.5] & 5322863/3526769 & 26.6093/17.6305       & $0.0197^{+0.0009}_{-0.0008}$/$0.0178^{+0.0007}_{-0.0006}$ & $1.1771^{+0.0220}_{-0.0198}$/$1.3297^{+0.0188}_{-0.0173}$ \\\hline\vspace*{1.5mm}
6& [0.2,0.4] & [-22.0,-21.0] & 5070090/3850127 & 25.3457/19.247       & $0.0151^{+0.0004}_{-0.0003}$/$0.0140^{+0.0003}_{-0.0003}$ & $1.2914^{+0.0109}_{-0.0103}$/$1.4010^{+0.0104}_{-0.0099}$ \\\hline\vspace*{1.5mm}
7& [0.2,0.4] & [-23.0,-22.0] & 610255/531282     & 3.0507/2.6559    & $0.0087^{+0.0002}_{-0.0002}$/$0.0080^{+0.0002}_{-0.0002}$ & $1.6975^{+0.0080}_{-0.0077}$/$1.7701^{+0.0070}_{-0.0070}$ \\\hline\vspace*{1.5mm}
8& [0.3,0.5] & [-21.0,-20.5] & 9482610/5793204 & 30.1603/18.4258        & $0.0265^{+0.0025}_{-0.0019}$/$0.0223^{+0.0012}_{-0.0011}$ & $1.2572^{+0.0505}_{-0.0406}$/$1.3740^{+0.0318}_{-0.0278}$ \\\hline\vspace*{1.5mm}
9& [0.3,0.5] & [-22.0,-21.0] & 8633198/6279861 & 27.4587/19.9737       & $0.0190^{+0.0006}_{-0.0006}$/$0.0172^{+0.0005}_{-0.0005}$ & $1.3494^{+0.0180}_{-0.0167}$/$1.4553^{+0.0157}_{-0.0146}$ \\\hline\vspace*{1.5mm}
10& [0.3,0.5] & [-23.0,-22.0] & 1082075/899869   & 3.4416/2.8621   & $0.0099^{+0.0002}_{-0.0002}$/$0.0090^{+0.0002}_{-0.0002}$ & $1.7283^{+0.0076}_{-0.0073}$/$1.8351^{+0.0075}_{-0.0073}$ \\\hline\vspace*{1.5mm}
11& [0.4,0.6] & [-21.5,-21.0] & 8313880/5541092 & 19.0616/12.7043       & $0.0213^{+0.0010}_{-0.0009}$/$0.0194^{+0.0007}_{-0.0006}$ & $1.3579^{+0.0268}_{-0.0239}$/$1.4873^{+0.0209}_{-0.0198}$ \\\hline\vspace*{1.5mm}
12& [0.4,0.6] & [-22.0,-21.5] & 3796681/2715952 & 8.7048/6.2270        & $0.0158^{+0.0004}_{-0.0004}$/$0.0147^{+0.0003}_{-0.0003}$ & $1.4767^{+0.0130}_{-0.0120}$/$1.6258^{+0.0129}_{-0.0121}$ \\\hline\vspace*{1.5mm}
13& [0.4,0.6] & [-23.0,-22.0] & 1813560/1355864  & 4.1580/3.1087        & $0.0099^{+0.0002}_{-0.0002}$/$0.0089^{+0.0002}_{-0.0002}$ & $1.7552^{+0.0071}_{-0.0069}$/$1.9276^{+0.0068}_{-0.0066}$ \\\hline\vspace*{1.5mm}
14& [0.5,0.7] & [-21.5,-21.0] & 10118646/6161085& 18.1256/11.0364      & $0.0221^{+0.0013}_{-0.0011}$/$0.0212^{+0.0011}_{-0.0009}$ & $1.3865^{+0.0343}_{-0.0299}$/$1.6075^{+0.0331}_{-0.0290}$ \\\hline\vspace*{1.5mm}
15& [0.5,0.7] & [-22.0,-21.5] & 5509779/3639334 & 9.8697/6.5192        & $0.0192^{+0.0008}_{-0.0007}$/$0.0187^{+0.0007}_{-0.0006}$ & $1.5620^{+0.0244}_{-0.0224}$/$1.7465^{+0.0241}_{-0.0220}$ \\\hline\vspace*{1.5mm}
16& [0.5,0.7] & [-23.0,-22.0] & 2915483/2002685  & 5.2225/3.5874       & $0.0129^{+0.0002}_{-0.0002}$/$0.0118^{+0.0002}_{-0.0002}$ & $1.7861^{+0.0099}_{-0.0095}$/$1.9824^{+0.0100}_{-0.0099}$ \\\hline\vspace*{1.5mm}
17& [0.6,0.8] & [-22.0,-21.5] & 7093064/4279621 & 10.4952/6.3323       & $0.0230^{+0.0014}_{-0.0012}$/$0.0232^{+0.0014}_{-0.0012}$ & $1.6575^{+0.0424}_{-0.0371}$/$1.8474^{+0.0471}_{-0.0419}$ \\\hline\vspace*{1.5mm}
18& [0.6,0.8] & [-23.0,-22.0] & 3846490/2478615 & 5.6914/3.6675       & $0.0178^{+0.0006}_{-0.0005}$/$0.0175^{+0.0006}_{-0.0005}$ & $1.9046^{+0.0227}_{-0.0213}$/$2.1385^{+0.0253}_{-0.0234}$ \\\hline\vspace*{1.5mm}
19& [0.7,0.9] & [-23.0,-22.0] & 5096945/3130438 & 6.4861/3.9836       & $0.0240^{+0.0014}_{-0.0012}$/$0.0244^{+0.0015}_{-0.0012}$ & $2.0244^{+0.0500}_{-0.0437}$/$2.3440^{+0.0591}_{-0.0500}$ \\\hline\vspace*{1.5mm}
20& [0.8,1.0] & [-23.0,-22.0] & 5854132/3323035 & 6.6040/3.7487       & $0.0261^{+0.0019}_{-0.0016}$/$0.0243^{+0.0014}_{-0.0012}$ & $1.9435^{+0.0616}_{-0.0537}$/$2.1950^{+0.0551}_{-0.0479}$ \\\hline
\bottomrule
\end{tabular*}
\end{table*}

\subsection{Galaxy Clustering Measurements}
\label{subsec:measurements}

In W19, we demonstrated that intrinsic clustering and photoz uncertainty can be 
simultaneously constrained using the projected 2PCF. 
In this work we measure the projected 2PCF through the following equation,
\begin{equation}\label{eq:wp}
    w_{p}^{\rm obs}(r_{\rm p}|r_{\pi,\max}) = 2\int_0^{r_{\pi, \max}}\xi^{\rm obs}(r_{p},r_{\pi})d{r_{\pi}}.
\end{equation}
where $r_p$ and $r_{\pi}$ are the transverse and line-of-sight separation
between galaxy pair, $r_{\pi, \max}$ is the  upper bound of the integration
along the line-of-sight direction. 
Since the methodology requires to
measure the projected 2PCF with two different $r_{\pi, \max}$, we choose
$r_{\pi, \max}=50$ and $100 h^{-1}\rm Mpc$ in this work for $w_{p}$ measurements
for all the galaxy (sub-)samples.
$\xi^{\rm obs}$ is the two-dimensional
correlation function in redshift space. We estimate $\xi^{\rm obs}(r_p,
r_{\pi})$ using the Landy-Szalay estimator     
\cite{correlation_function, 1998ApJ...494L..41S, 2000ApJ...535L..13K},
\begin{equation}\label{eq:xi}
    \xi^{\rm obs}(r_{\rm p},r_{\pi}) = \frac{\rm DD-2DR +RR}{\rm RR}
\end{equation}
where ${\rm DD}$, ${\rm RR}$ and ${\rm DR}$ are normalized number of galaxy-galaxy, 
random-random and galaxy-random pairs within the projected separation $r_p$ and the line-of-sight separation $r_{\pi}$, respectively.

We use the public random catalogs ({\it randoms-inside \footnote{\url{https://www.legacysurvey.org/dr8/files/\#random-catalogs}}})
released along with LS DR8 to account 
for the survey geometry and angular selection function. When making the 
random catalogs, we adopt the same cut (i.e., require objects to have at 
least 1 exposure in each optical band, remove low galactic altitude region,
and apply same masks) 
to make sure that random catalog have the same angular selection with the corresponding galaxy sample. Here we use the  `shuffled' redshift method to generate the random samples to compensate possible systematic selection effects as a function of redshift in the observational (sub-)samples.
The random catalog for each galaxy (sub-)sample contains $\sim$5 times as many galaxies.
We employ the python package {\it CorrFunc} to compute the pair-counts
\cite{10.1007/978-981-13-7729-7_1,2020MNRAS.491.3022S}.

We adopt the sample mean from jackknife resampling method 
as our estimation of projected 2PCF.
The covariance matrix is also estimated with the jackknife method. 
The footprint of the galaxy sample is divided into $N = 120$ spatially contiguous and equal area sub-regions utilizing the python package 
{\it kmeans\_radec \footnote{\url{https://github.com/esheldon/kmeans_radec}}}. 
We measure $w_{p}$ for 120 times leaving out one different sub-region 
at each time, and the covariance is calculated as 119 times the variance 
of the 120 measurements 
\cite{2005ApJ...630....1Z,zehavi11,guo15,2016MNRAS.460.3647X,2016MNRAS.457.4360Z,2017ApJ...846...61G}.

\begin{figure*}[!t]
\centering
\includegraphics[height=0.8\textwidth,width=0.95\textwidth]{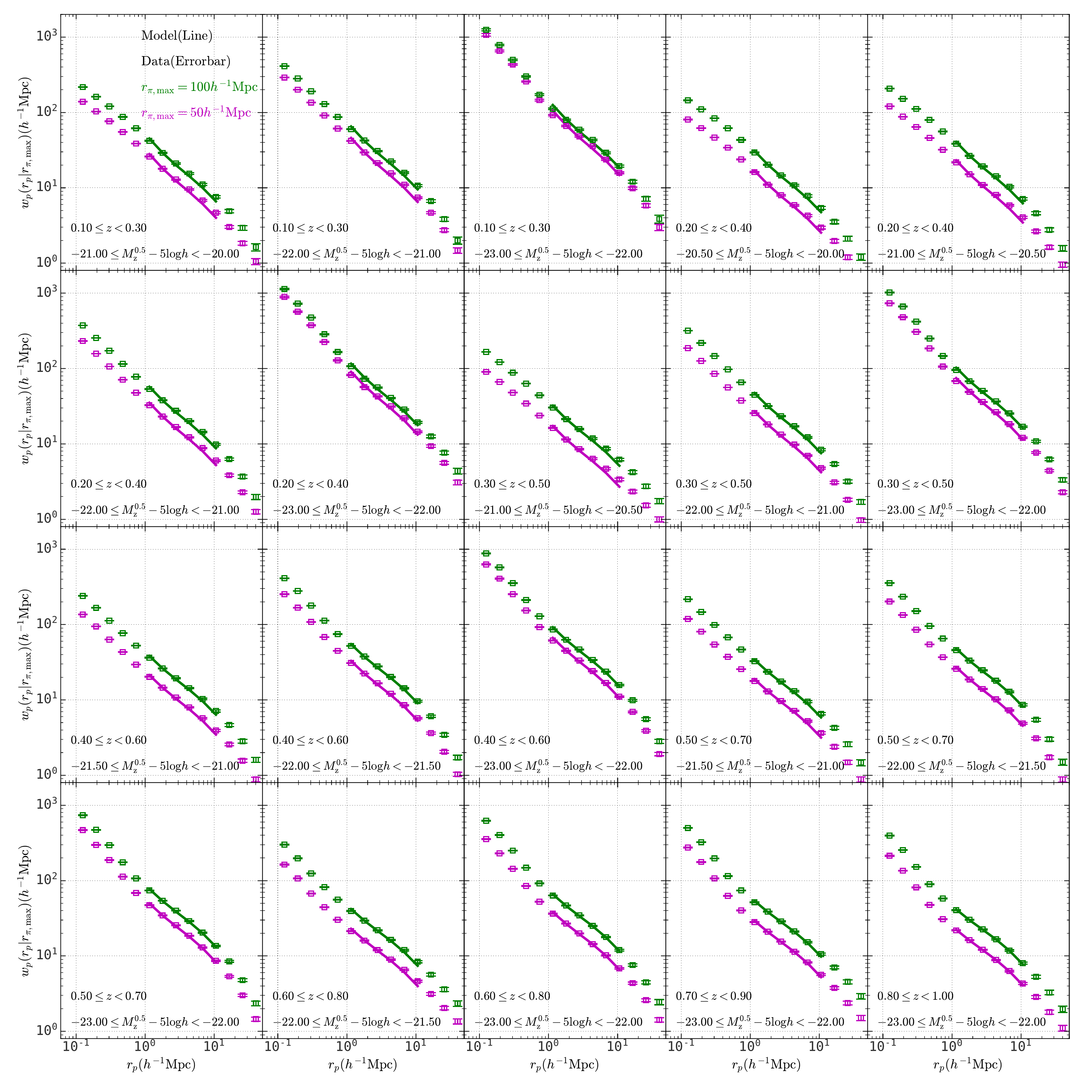}
    \caption{Observational measurements and model predictions of the projected 2PCFs for two different
        $r_{\pi,\max} = 50~(100)\mpch$ values. We use open squares with
        errorbars to represent the observational measurements 
        and solid lines to represent the model predictions. Shown in different (in total 20) panels are results for
        different galaxy samples (as indicated). }
    \label{fig:corr}
\end{figure*}

\begin{figure*}[!t]
\centering
\includegraphics[height=0.8\textwidth,width=0.95\textwidth]{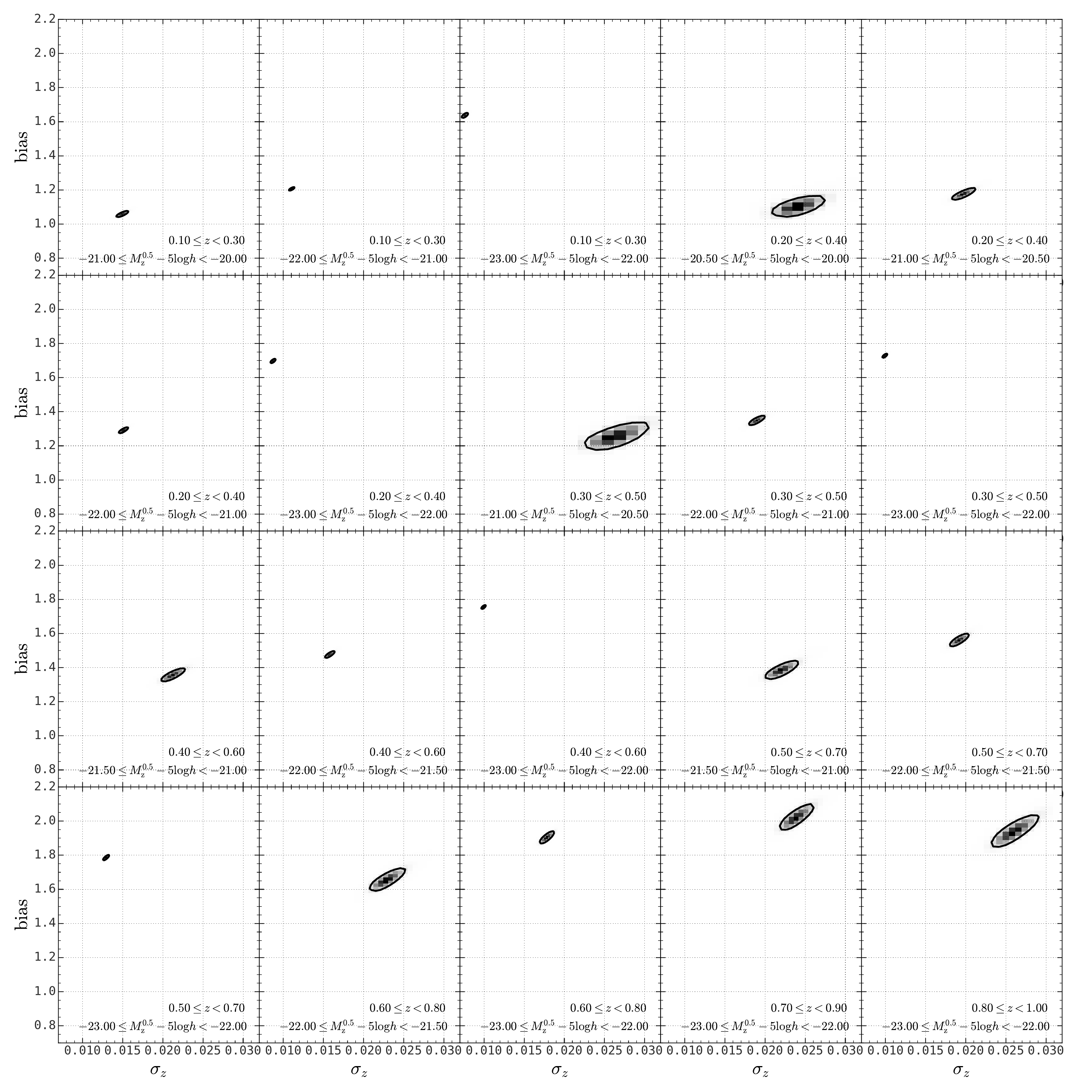}
    \caption{The 2D contours of the parameters $\sigma_{z}$ and ${\rm
        bias}$ for 20 different all galaxy samples. The black solid contour in each panel
    corresponds to the 68\% confidence region.}\label{fig:sigma}
\end{figure*}

\section{Methodology}\label{sec:method}

The method and performance tests of our photometric redshift modeling of the 2PCFs  were presented in W19 using N-body simulations, here we just briefly summarize
the key points.

We model the photometric redshift uncertainty as a Gaussian distribution, which
is
\begin{align}
    P\left(z_{\rm phot} - z_{\rm spec}\right) = \mathcal{N}(0, \sigma_z(1 + z_{\rm spec}))
\end{align}
where $z_{\rm phot}$ is the photometric redshift, $z_{\rm spec}$ is the
spectroscopic redshift, $\mathcal{N}(\mu, \sigma)$ is a Gaussian distribution
with mean $\mu$ and scale $\sigma$, and $\sigma_z$ is the uncertainty of the
photometric redshift we want to infer.

To calculate the 2PCF, we need to convert the redshift to comoving distance,
which is
\begin{align}
    D_c(z_{\rm phot}) &= \frac{c}{H_0}\int_0^{z_{\rm phot}}\frac{dz^{\prime}}{E(z^{\prime})}\\
                      &\approx D_c(z_{\rm spec}) + \frac{c(z_{\rm phot} - z_{\rm spec})}{H_0E(z_{\rm spec})}
\end{align}
where $E(z)=\sqrt{\Omega_{m}(1+z)^{3}+\Omega_{\Lambda}}$. In the 2PCF measurements, we count galaxy pairs as a function of their separations $DD(r_p,r_{\pi})$, where $r_{\pi}=D_c(z_{\rm 1}) - D_c(z_{\rm 2})$. Because of the photoz error, the probability
distribution of  the difference between photoz derived separation and spectroscopic redshift derived separation follows,
\begin{align}
    P(R) &= \frac{1}{\sqrt{2\pi}\sigma_{\rm R}}\exp\left(-\frac{R^2}{2\sigma_{\rm R}^2}\right)\\
    \sigma_R &= \frac{\sqrt{2}c\sigma_z(1 + z_{\rm spec})}{H_0 E(z_{\rm spec})}
\end{align}
where $R\equiv r_{\pi}(z_{\rm phot}) - r_{\pi}(z_{\rm spec})$.

Theoretically, the 2PCF of galaxies on large scales in redshift space can be described by, 
 \begin{equation}\label{eq:prud}
     \xi^{\rm model}(r_{\rm p},r_{\pi}) = \int^{\infty}_{-\infty}b^2\times\xi^{\rm mm}(r_{\rm p},r_{\pi}-R)P(R)dR
 \end{equation}
 where $\xi^{\rm mm}$ is the correlation function of matter calculated from
 CAMB \cite{CAMB}, and $b$ is the galaxy linear bias\footnote{More accurate modeling of the galaxy 2PCFs, especially on small one-halo term scales, can be achieved using the HOD/CLF models,  which we will present in a subsequent probe.}.  The projected 2PCF in the model
 can also be calculated as
 \begin{equation}\label{eq:wp}
     w_{\rm p}^{\rm model}(r_{\rm p}|r_{\pi,\max}) = 2\int_0^{\rm r_{\pi,
     \max}}\xi^{\rm model}(r_{\rm p},r_{\rm \pi})d{r_{\rm \pi}}.
 \end{equation}

Here we constrain the two free parameters, $\sigma_z$ and $b$, by maximizing
the posterior distribution
\begin{equation}
    P_{\rm posterior}(\sigma_z, b) = P_{\rm prior}(\sigma_z, b)\times \mathcal L(\mathcal D|\sigma_z, b)
\end{equation}
where the likelihood is
\begin{equation}\label{13}
    \log\mathcal L \propto ({\mathbf w_{p}^{\rm obs}}-{\mathbf w_{p}^{\rm model}})^{\rm T}
    {\mathbf C^{-1}}{({\mathbf w_{p}^{\rm obs}}-{\mathbf w_{p}^{\rm model}})}
\end{equation}
where $\mathbf w^{obs}_{p}$ is the concatenated vector of projected 2PCF integrated
to different upper bound, i.e. $\mathbf w^{obs}_{p} =[\mathbf w^{obs}_{p}(50h^{-1}{\rm
Mpc}), \mathbf w^{obs}_{p}(100h^{-1}{\rm Mpc})]$,
and ${\mathbf C}$ is the error covariance matrix of the data vector inferred
from the jackknife method \cite{2009MNRAS.396...19N, 2013ApJ...767..122G,
guo14}. The prior distributions for $\sigma_z$ and $b$ are uniform
distribution, and in the range of $[0.0001, 0.5\times
z_{\max}]$ and $[0.01, 10.0]$, respectively, where $z_{\max}$ is the maximum redshift in each
sample, because
$\sigma_{z}$ increases as redshift get higher. Here we employed the \texttt{emcee} \cite{emcee} package to sample the
posterior distribution of $\sigma_z$ and $b$.

\begin{figure*}[!t]
\centering
\includegraphics[height=0.37\textwidth,width=0.95\textwidth]{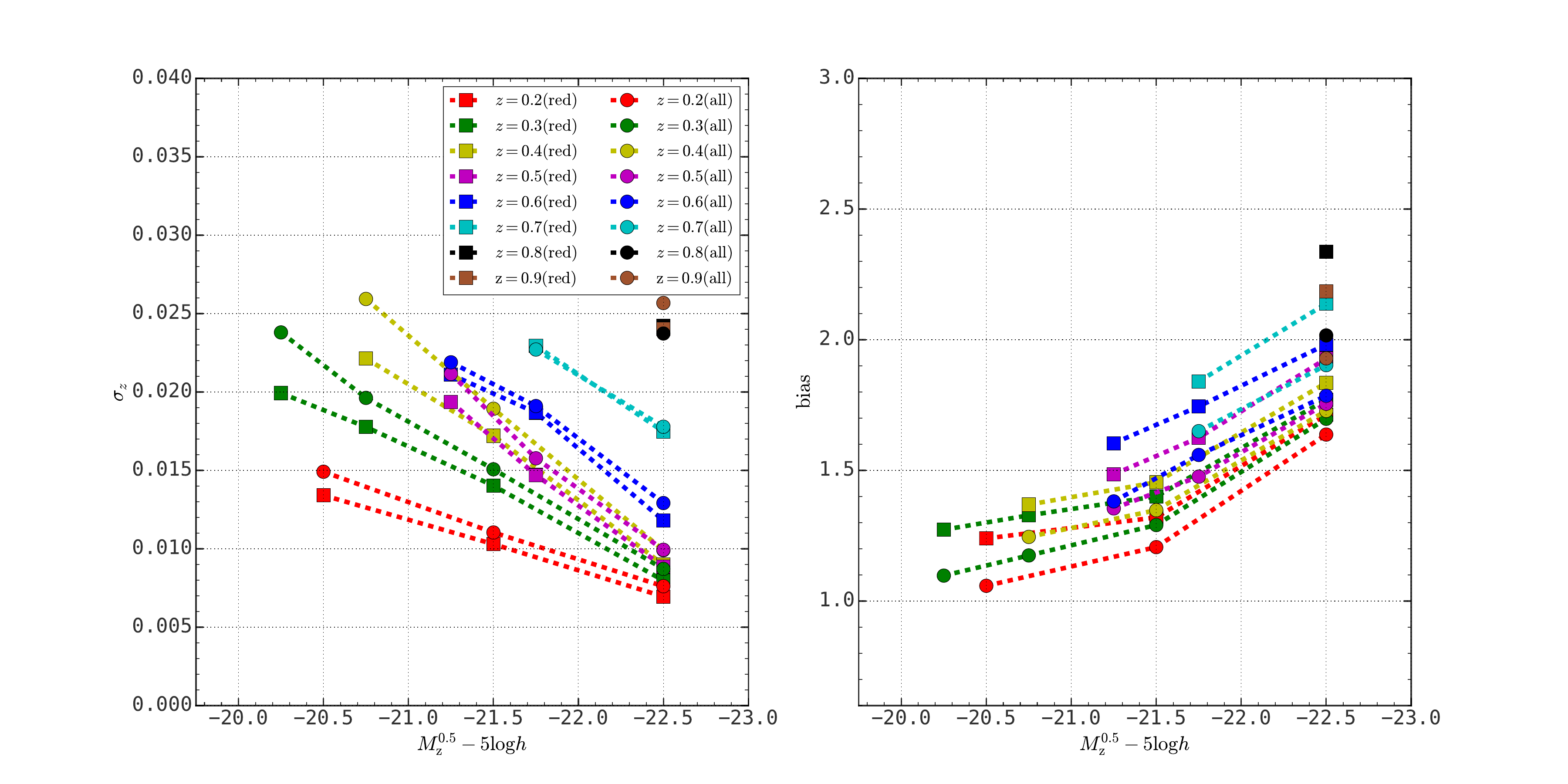}
    \caption{The best fit $\sigma_{z}$ and 
        bias measurements for galaxy (sub-)samples in the DESI Imaging Legacy Surveys within different absolute
        magnitude and redshift bins. Different colors represent galaxy  (sub-)samples in different
    redshifts bins. }
    \label{fig:sigma_bias}
\end{figure*}

\begin{figure*}[!t]
\centering
\includegraphics[height=0.8\textwidth,width=0.95\textwidth]{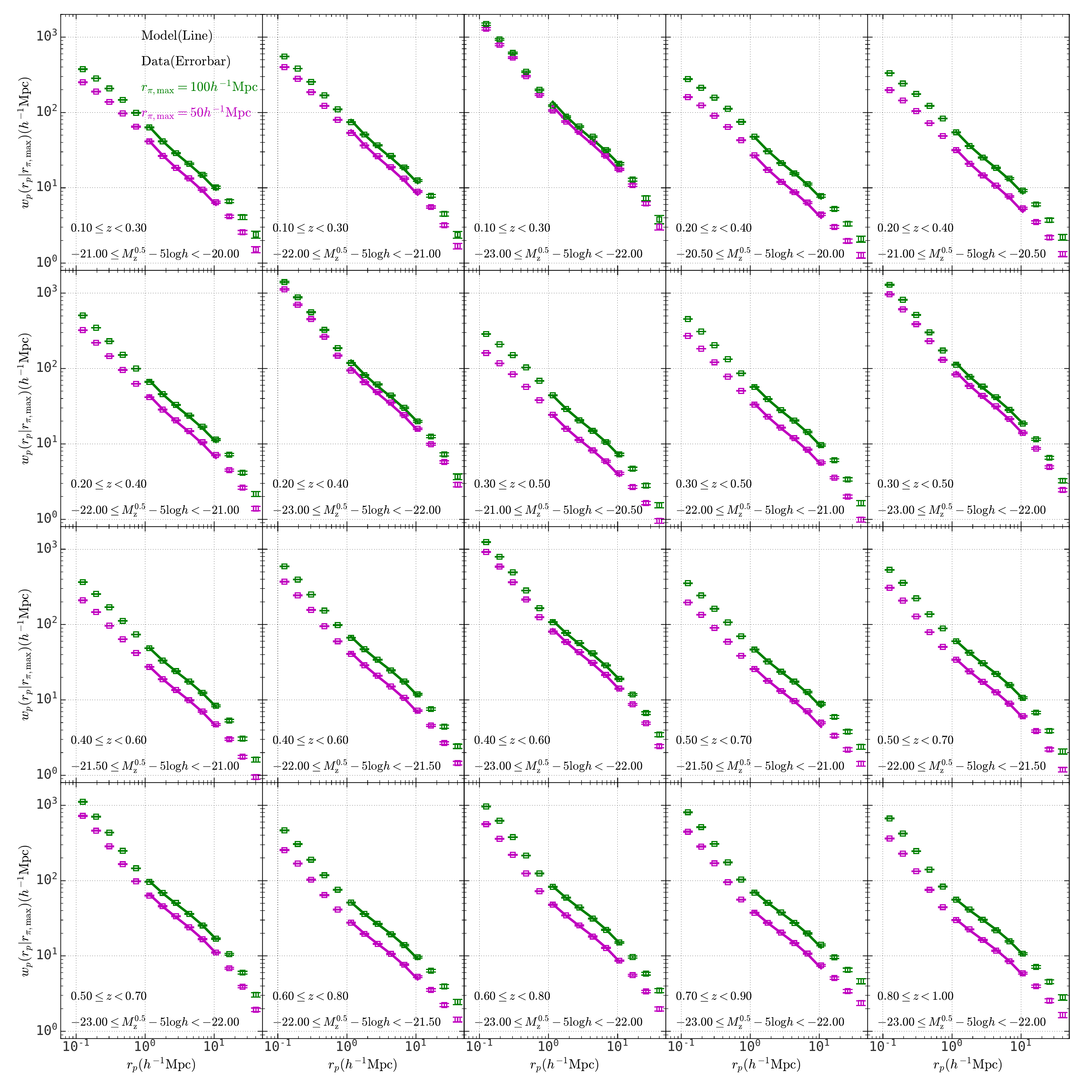}
    \caption{Similar to Fig. \ref{fig:corr}, but here for 20 red galaxy sub-samples.  }\label{fig:corr_red}
\end{figure*}

\begin{figure*}[!t]
\centering
\includegraphics[height=0.8\textwidth,width=0.95\textwidth]{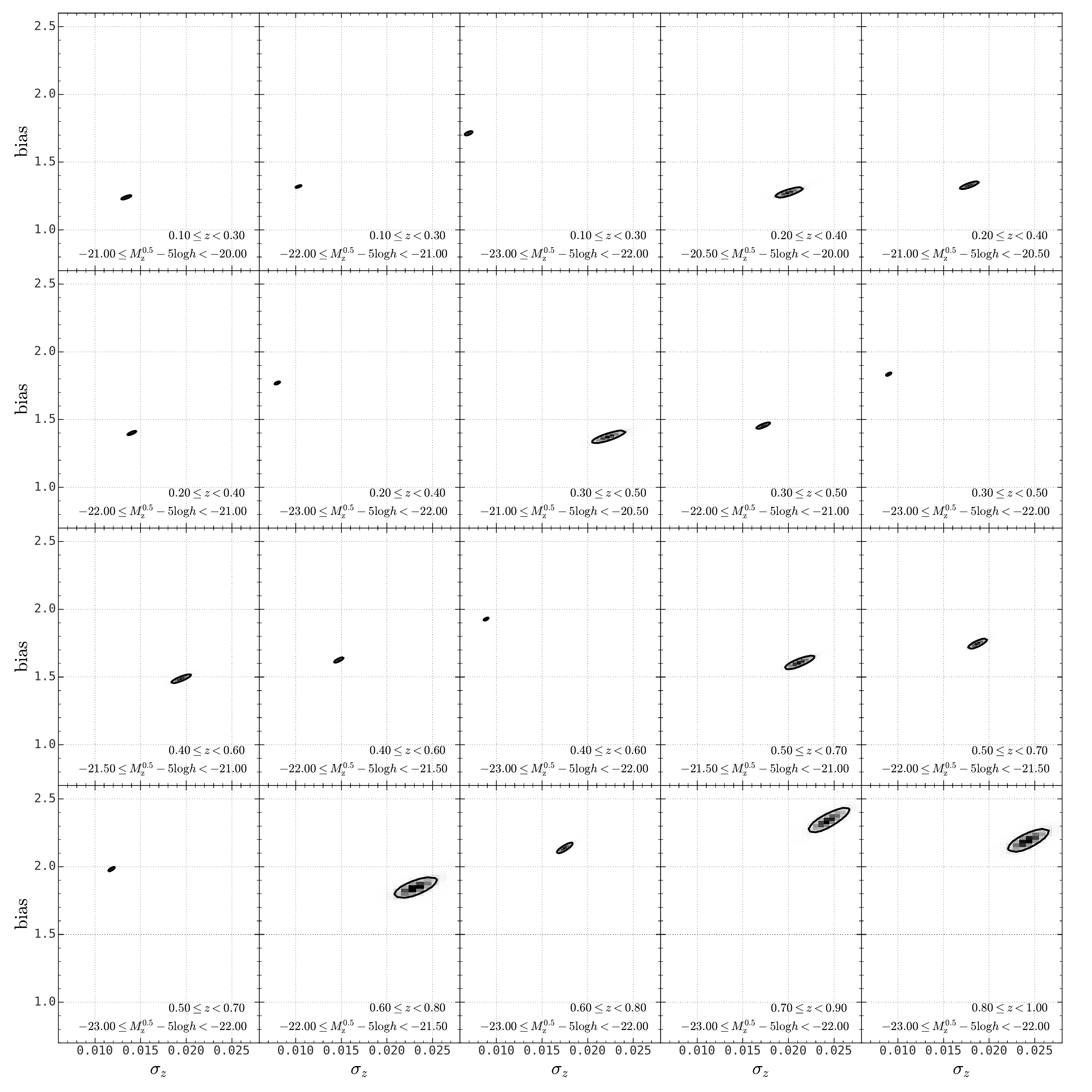}
    \caption{ Similar to Fig. \ref{fig:sigma}, but here for 20 red galaxy sub-samples. }\label{fig:sigma_red}
\end{figure*}

\section{Results}\label{sec:results}

\subsection{Model fitting}
\label{sec:mf}

We start our investigation with 20 all galaxy samples. In \cref{fig:corr}, we show $w^{\rm obs}_{p}$ measurements (open squares)
in different redshift and magnitude bins, where $r_{p}$ is from $10^{-1}$ to 
$10^{1.7}h^{-1}\rm Mpc$ in 15 logarithmic bins. Here results are shown for the projected 2PCFs calculated with  two different $r_{\pi,\max} = 50/100\mpch$ values. As shown in the figure, galaxies in brighter magnitude bins in general have stronger clustering strengths. 
While the clustering amplitude differences between two different $r_{\pi,\max}$ become closer as galaxies become brighter or in the lower redshift bins, which indicate relatively smaller $\sigma_z$.

Quantitatively, we use  $w^{\rm obs}_p(r_p|r_{\pi,\max})$ on scales from $r_p = 1$ to 10 $h^{-1}\rm Mpc$ to constrain our model parameters, $\sigma_z$ and $b$. The reasons of using these scales are two folded: (1) on smaller scales, the galaxy bias becomes nonlinear. (2) on larger scales, the 2PCFs may somehow suffer from the systematic photometric redshift errors.

For each galaxy sample, we run 10 Markov Chain Monte Carlo (MCMC) chains, each with 50,000 steps. The chains converge within the first 10,000
steps mostly, these steps are discarded because they are regarded as burn-in
stage. Finally we obtain about 400,000 MCMC models. The posterior distributions
of the model parameters are shown in \cref{fig:sigma}. The best-fit $\sigma_{z}$ and galaxy bias are listed in Table \ref{table:sample}. 
We show in Fig. \ref{fig:corr} using solid lines the best fitting model 
predictions. Overall, our model can accurately
recover the clustering signals in different redshift and luminosity bins.

Looking into the best fitting parameters themselves, as shown in the left panel of \cref{fig:sigma_bias} using open circles, we found the photoz errors get larger when galaxies become fainter and locate at larger redshifts. Quite interestingly, our model constraints on this value 
are in nice agreement with the direct measurements of the photoz errors
from a small sample of galaxies that have spectroscopic redshifts
\cite{2020arXiv200106018Z}.
The open circles shown in the right panel of \cref{fig:sigma_bias} are the best fit galaxy biases in different absolute magnitude and
redshift bins. Obviously, galaxy biases become larger when galaxies are brighter and locate at higher redshifts. For the brightest galaxy samples, the bias increase from $\sim 1.6$ at $z\sim 0.2$ to $\sim 2.0$ at $z\sim 0.8$.

\begin{figure}[H]
\centering
\includegraphics[height=0.45\textwidth,width=0.47\textwidth]{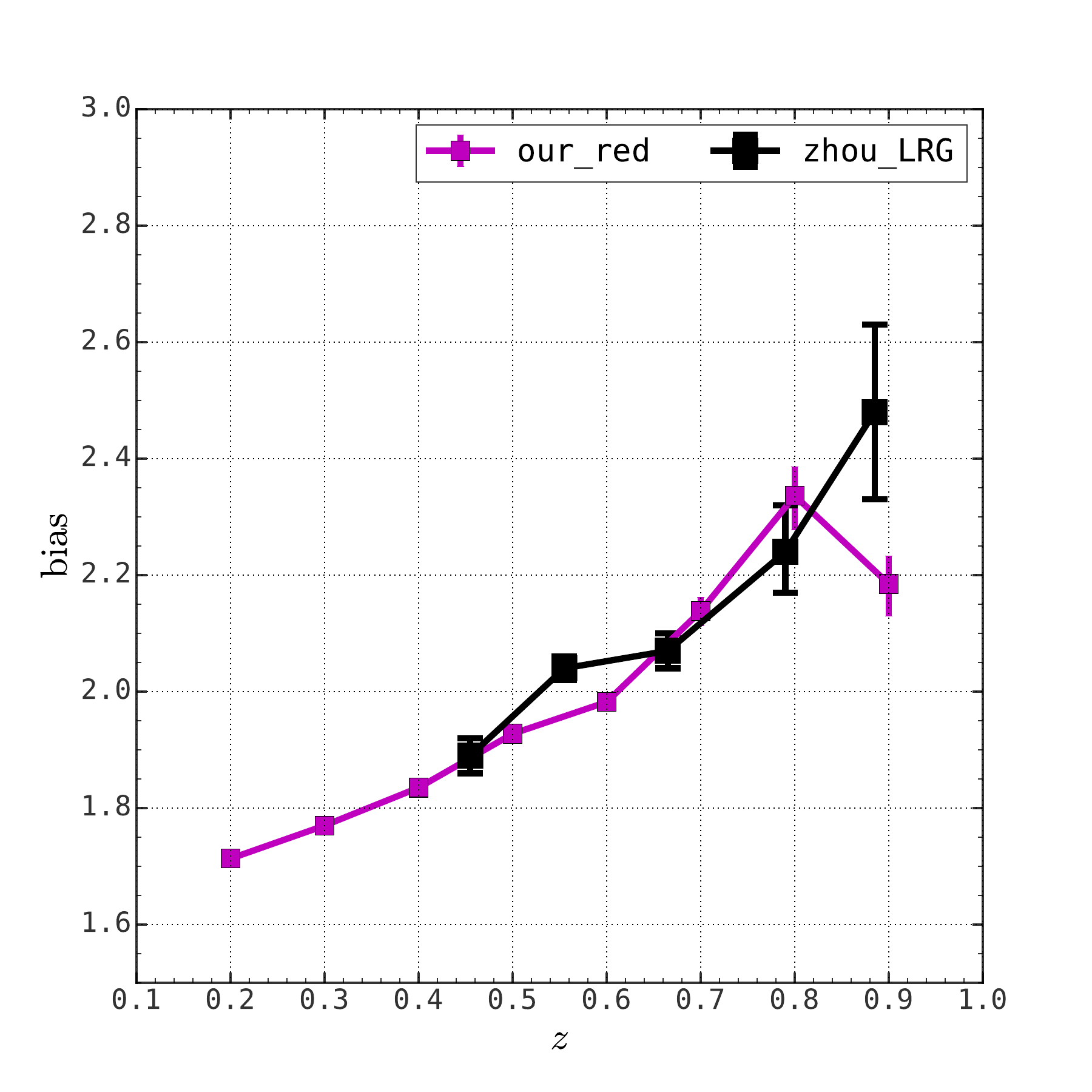}
    \caption{Our best fit bias measurements for the brightest red galaxy sub-samples in the DESI Imaging Legacy Surveys compared with the bias measurements from luminous red galaxy (LRG) samples obtained by \cite{2020arXiv200106018Z} as a function of  redshift. }\label{fig:bias_zhou}
\end{figure}

\begin{figure*}[!t]
\centering
\includegraphics[height=0.8\textwidth,width=0.95\textwidth]{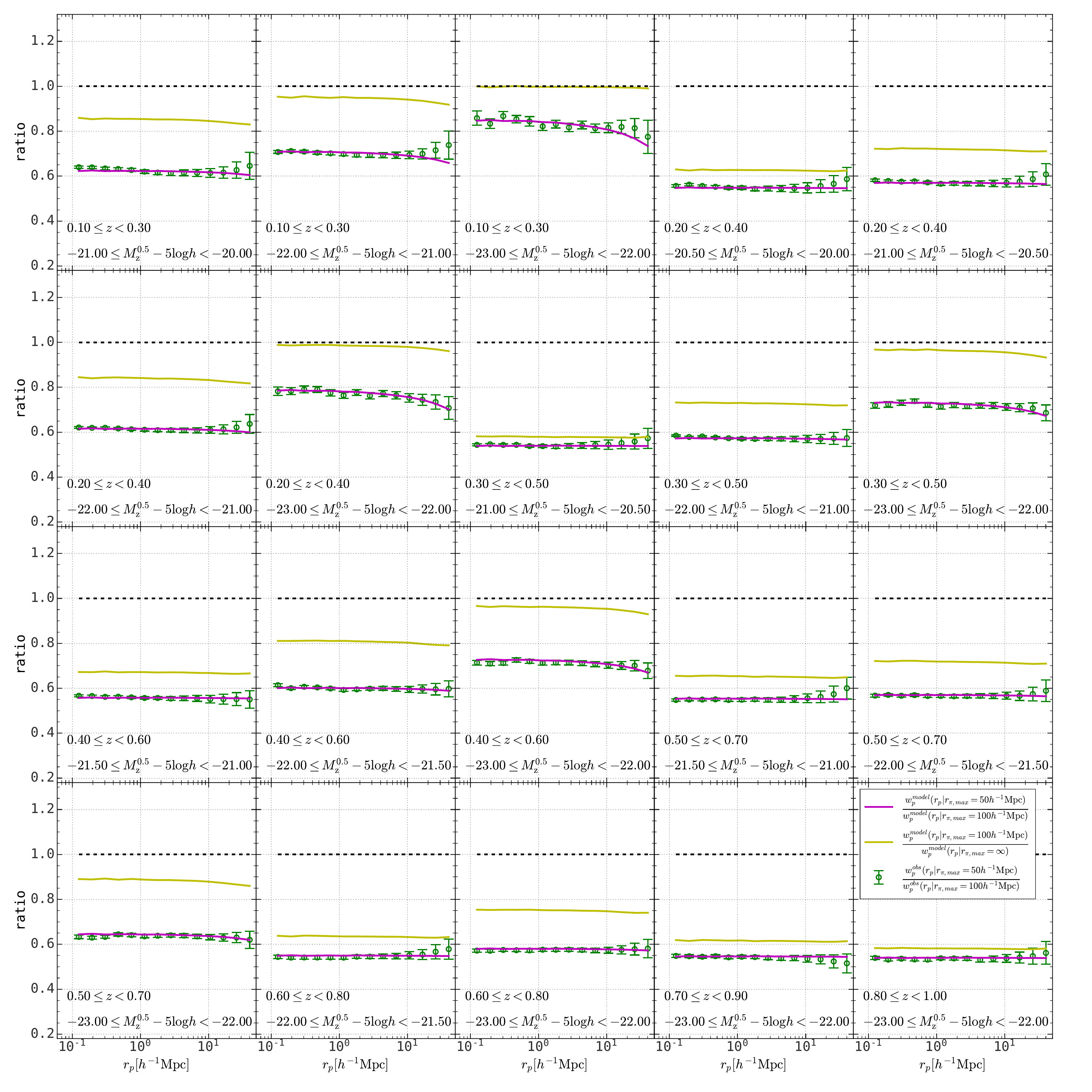}
    \caption{The ratio of projected 2PCFs. The purple lines and open circles with error-bars are results for the model $w_{p}^{\rm model}(r_{\rm p}|r_{\pi,\max})$  ratios and the observed $w_{p}^{\rm obs}(r_{\rm p}|r_{\pi,\max})$ ratios for the two $r_{\pi,\max} = 50$ and $100\mpch$ cuts, respectively. The yellow solid lines are  model predictions of $w_{p}^{\rm model}(r_{\rm p}|r_{\pi,\max})/w_{p}^{\rm model}(r_{\rm p})$ for $r_{\pi,\max} = 100\mpch$ cut. }\label{fig:ratio}
\end{figure*}

\begin{figure*}[!t]
\centering
\includegraphics[height=0.8\textwidth,width=0.95\textwidth]{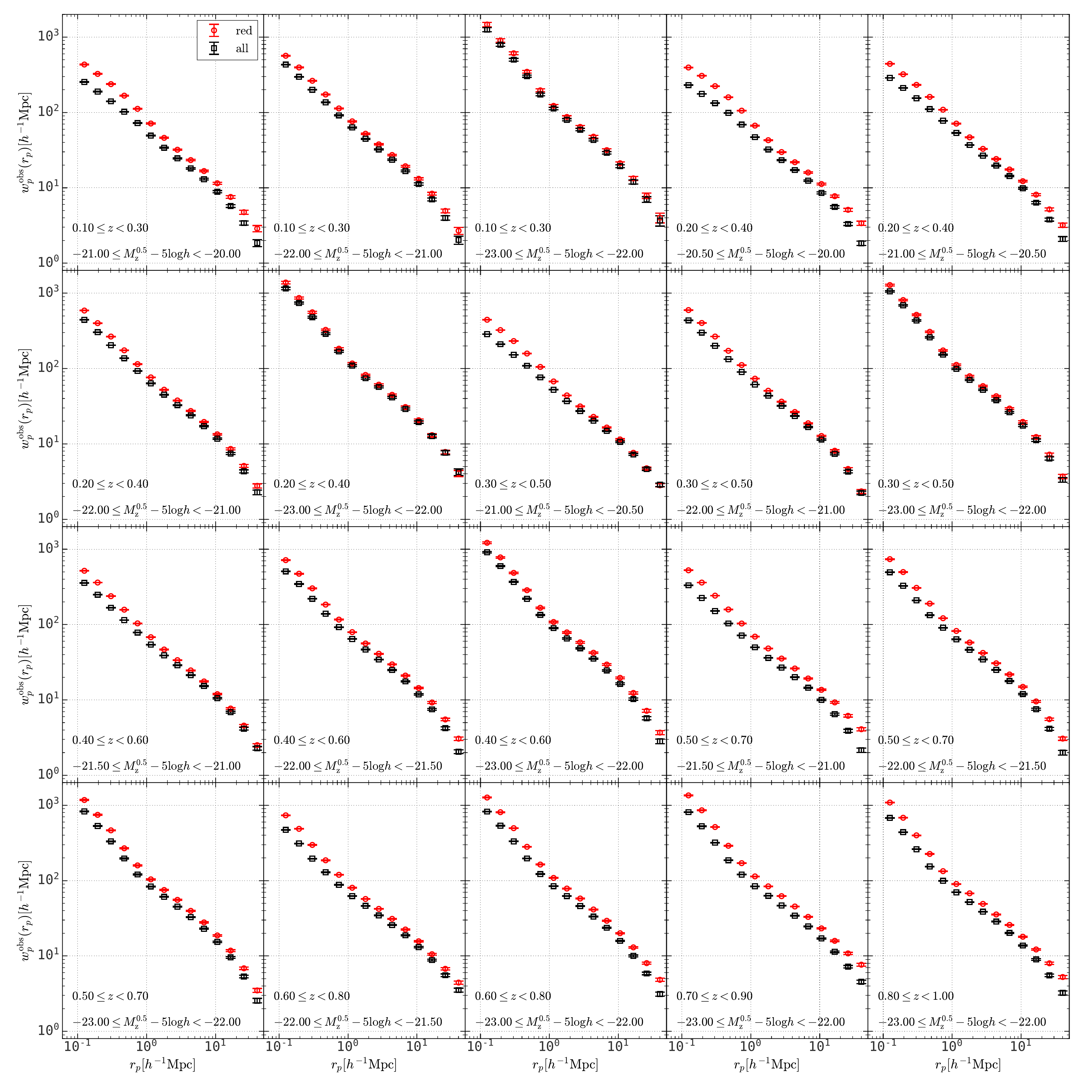}
    \caption{The extracted intrinsic projected 2PCFs for all and red galaxy (sub-)samples. Here the error-bars in each (sub-)sample are enlarged by a factor of 2 with respect to the direct measurements in $w_{p}^{\rm obs}(r_{\rm p}|r_{\pi,\max})$ with $r_{\pi,\max} = 100\mpch$ cut. }\label{fig:wp_intri}
\end{figure*}

Next, we turn to the 20 red galaxy sub-samples. Similar to the 20 all galaxy samples, we also measure their $w^{\rm obs}_{p}$ with two different
$r_{\pi,\max} = 50/100\mpch$ values. The results are shown in Fig. \ref{fig:corr_red} using open squares.  The overall behaviors are quite similar to those of all galaxy samples. Here we
can see that the gap between $w^{\rm obs}_{p}$ for different $r_{\pi,\max}$ values become even smaller in these red sub-samples.
 
Following the same procedures for each of the 20 all galaxy samples, we run 10 MCMC chains, each with 50,000 steps for each of our red galaxy subsamples. After discarding the  burn-in  stage models, we use the remaining 400,000 MCMC models to describe the  posterior distributions of the model parameters. The results  are shown in Fig. \ref{fig:sigma_red} with contours. We also list the best-fit $\sigma_{z}$ and galaxy bias for our 20 red galaxy subsamples in Table \ref{table:sample}.  As an illustration, we show in Fig. \ref{fig:corr_red} using solid lines the best fitting model 
predictions. Here again, our models accurately
recover the clustering signals in different redshift and luminosity bins.  

In addition to the red galaxy sub-samples, we also looked into the 20 blue galaxy sub-samples.
However, as we will demonstrate in the Appendix, the blue sub-samples might suffer from systematic 
photometric redshift uncertainties, where their clustering measurements can not be well modelled. 

Because of these, we only focus on the red and all galaxy (sub-)samples to model their color dependence. As shown in the Fig. \ref{fig:sigma_bias}, red galaxy sub-samples show smaller $\sigma_{z}$ and higher galaxy bias in the same redshift and absolute magnitude bins. The difference is larger in less bright galaxy (sub-)samples.

Finally, we compare our bias measurements with the ones obtained in a recent similar study  \cite{2020arXiv200106018Z} in Fig. \ref{fig:bias_zhou}. 
The black squares with error bars shown in this figure are the bias measurements obtained by Zhou et al. \cite{2020arXiv200106018Z} from the DECaLS DR7 using an HOD fitting algorithm,  while our results are shown as the magenta squares. In that study, they also  used the projected 2PCFs measured from the photometric redshift data  and made their model constraints. However, unlike our two integration depths method, they 
treat photoz error as a prior to assign a redshift error to each galaxy in their mock samples. By comparing the observed 2PCFs with the mock 2PCFs they fitted the related HOD model parameters,  and hence obtained the biases of galaxies. To have fair comparisons with their  luminous red galaxy (LRG) samples, we only show our best fit bias measurements for the brightest red galaxy sub-samples. Taking into account the different galaxy number densities between the two sets of galaxy samples, overall, our results agree with their measurements quite well at redshift $z> 0.6$. Indeed the discrepancy in the highest redshift bin is mainly induced by the galaxy selection where their galaxy number density in that bin is much lower than ours. On the other hand, their biases measured from the two lower redshift bins are somewhat larger than ours. We argue that this might be caused by a bit too large assigned photoz errors in their modelling. Note that our measurements and model constraints show that photoz errors decrease significantly at lower redshifts.

\subsection{The intrinsic projected 2PCFs}

After we modeled the projected 2PCFs to different $r_{\pi, \max}$ using the $\sigma_z$ and bias parameters, we can in general get an estimate of their intrinsic values, i.e. the projected 2PCFs in real space, which corresponds to $w_{p}^{\rm obs}(r_{\rm p}) = w_{p}^{\rm obs}(r_{\rm p}|r_{\pi,\max}=\infty)$ here.  This set of measurements are very important for galaxy formation and cosmological 
  constraints in the HOD modelings. We use the following equation
to estimate the observational intrinsic projected 2PCFs,
\begin{equation}\label{eq:wpobs}
    w_{p}^{\rm obs}(r_{\rm p}) = w_{p}^{\rm obs}(r_{\rm p}|r_{\pi,\max})*w_{p}^{\rm model}(r_{\rm p})/w_{p}^{\rm model}(r_{\rm p}|r_{\pi,\max})\,,
\end{equation}
where we use  $r_{\pi,\max} = 100\mpch$ cut to make the estimation. 
Note that this equation holds accurately if the observed 2PCFs are related with the dark matter 2PCF by
a constant bias factor. Unless the bias factor has a very strong scale dependence,  especially on small scales,
the correction factor $w_{p}^{\rm model}(r_{\rm p})/w_{p}^{\rm model}(r_{\rm p}|r_{\pi,\max})$ may be slightly
different for the observed galaxies.  

To test the reliability and assess the accuracy of using the model $w_{p}^{\rm model}(r_{\rm p})/w_{p}^{\rm model}(r_{\rm p}|r_{\pi,\max})$ ratio
based on the auto correlation function of dark matter particles to account for that of galaxies, we show in Fig. \ref{fig:ratio} for 20 all galaxy samples a comparison of the model $w_{p}^{\rm model}(r_{\rm p}|r_{\pi,\max})$  ratios and the observed $w_{p}^{\rm obs}(r_{\rm p}|r_{\pi,\max})$ ratios for the two $r_{\pi,\max} = 50$ and $100\mpch$ cuts, respectively. 
The open circles with error-bars shown in Fig. \ref{fig:ratio} are the results for the observed $w_{p}^{\rm obs}(r_{\rm p}|r_{\pi,\max})$ ratios, which show very weak scale dependence. The purple solid lines are the  results for the  model $w_{p}^{\rm model}(r_{\rm p}|r_{\pi,\max})$ ratios. Note that although our model parameters are only constrained using the $w_{p}^{\rm obs}(r_{\rm p}|r_{\pi,\max})$ values on scales between  $r_p = 1$ and 10 $h^{-1}\rm Mpc$, overall, the predicted model ratios agree with
the observed ratios very well, i.e., within 1-$\sigma$ error-bars on all the scales. Next, the model predictions of $w_{p}^{\rm model}(r_{\rm p}|r_{\pi,\max})/w_{p}^{\rm model}(r_{\rm p})$ for $r_{\pi,\max} = 100\mpch$ cut are shown in Fig. \ref{fig:ratio} using yellow solid lines. In all the cases, we find that this set of correction factors
are all larger (i.e., less significant) than those between $r_{\pi,\max} = 50$ and $100\mpch$ cuts. And the scale dependence is also weaker. Based on these features, we infer that the overall 
correction in Eq. \ref{eq:wpobs} is reliable and the errors induced in this step should not exceed 1-$\sigma$ errors of the observational data. 
To be conservative, we will enlarge the error-bars in our extracted intrinsic projected 
2PCFs by a factor of two for future use. We also checked the situation for the 20 red 
galaxy sub-samples, for simplicity not explicitly shown here, the overall behaviors are
quite similar to the 20 all galaxy samples.

Shown in Fig. \ref{fig:wp_intri} are the intrinsic projected 2PCFs we obtained for the all and red
galaxy (sub-)samples. As we mentioned, here the error-bars in each (sub-)sample are enlarged by a factor of 2 with respect to the direct measurements in $w_{p}^{\rm obs}(r_{\rm p}|r_{\pi,\max})$ with $r_{\pi,\max} = 100\mpch$ cut. We can see that overall, in the same redshift and luminosity bin, the red galaxies show somewhat stronger clustering strength than the all galaxies. The projected 2PCFs on small scales is more enhanced in red sub-samples than the all samples. These features in general hold important information regarding how galaxies with different colors populated dark matter halos and the related galaxy formation 
and evolution processes. We will come to this topic in subsequent probes.



\section{Summary}\label{sec:summary}

In this study, we construct 60 volume-limited galaxy (sub-)samples, sampling 
8 redshift bins from $z=0.1$ to $z=1.0$, and a few $\rm z$-band absolute magnitude 
bins with $M_{\rm z} \le -20$,  from the photometric redshift galaxy catalogs of the  
DESI Imaging Legacy Surveys DR8. We measure the projected 2PCFs for all these 
60 (sub-)samples with two  $r_{\pi,\max} = 50$ and $100\mpch$ cuts along the 
line-of-sight directions, respectively. 
Using the photometric redshift modeling of the 2PCFs developed in W19, we 
constrain the photoz errors and galaxy biases for all the 60 volume-limited 
galaxy (sub-)samples. Our main results are summarized as follows.
\begin{itemize}
\item Our model can well describe the clustering properties of the red and all
galaxy (sub-)samples, while the blue galaxy sub-samples might suffer from the 
systematic redshift errors, especially for low redshift bins, where the clustering 
on large scales are significantly enhanced and can not be described by the theoretical models.

\item Focusing only on the 40 red and all galaxy (sub-)samples, we find galaxies 
show better photoz performance and have higher biases when they become redder, 
brighter or in a lower redshift bin.

\item Based on the projected 2PCFs for the $r_{\pi,\max} = 100\mpch$ cut, and the 
theoretical model prediction of the correction factor, we obtain the intrinsic 
projected 2PCFs for 40 red and all galaxy (sub-)samples.
\end{itemize}

Here we note that we did not take
photoz redshift outliers into account, which according to \cite{2020arXiv200106018Z} are at less than 1\% level. 
And we assumed that the photometric redshift errors follow Gaussian distribution which is also quite well demonstrated in \cite{2020arXiv200106018Z}. With the intrinsic clustering measurements for 40 red and all galaxy (sub-)samples, we will go deep into
HOD/CLF framework to explore the halo-galaxy connection in a subsequent probe.

\Acknowledgements{This work is supported by the
    national science foundation of China (grant Nos. 11833005, 11890691,
    11890692, 11533006, 11621303), Shanghai Natural Science Foundation,
    Grant No. 15ZR1446700 and 111 project No. B20019. The
    Photometric Redshifts for the Legacy Surveys (PRLS) catalog used in this
    paper was produced thanks to funding from the U.S. Department of Energy
Office of Science,  Office of High Energy Physics via grant DE-SC0007914.}

\InterestConflict{The authors declare that they have no conflict of interest.}



\begin{appendix}




\renewcommand{\thesection}{Appendix}

\section{}

\begin{figure*}[!t]
\centering
\includegraphics[height=0.8\textwidth,width=0.95\textwidth]{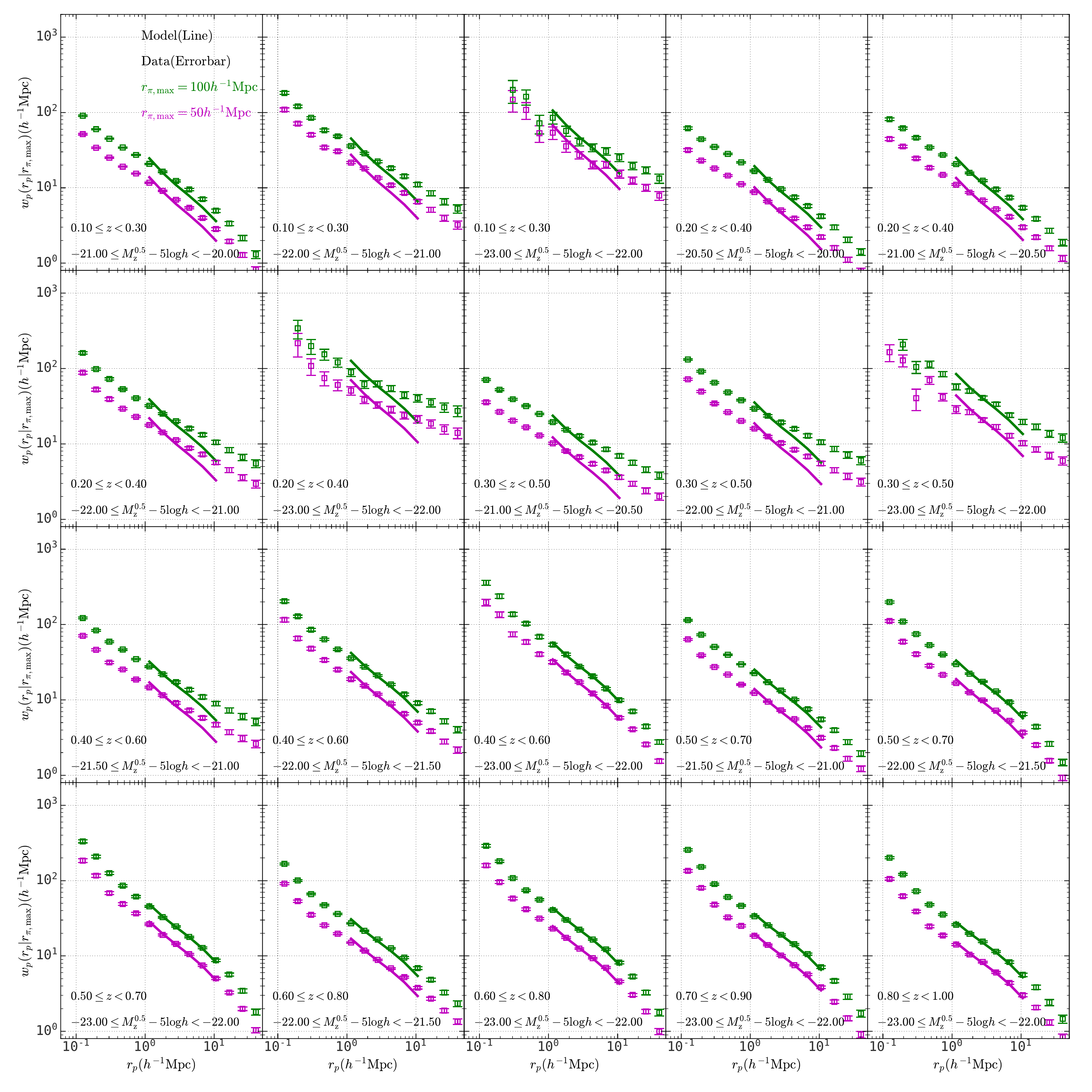}
    \caption{Similar to Fig. \ref{fig:corr}, but here for 20 blue galaxy sub-samples. }\label{fig:corr_blue}
\end{figure*}

\begin{figure*}[!t]
\centering
\includegraphics[height=0.8\textwidth,width=0.95\textwidth]{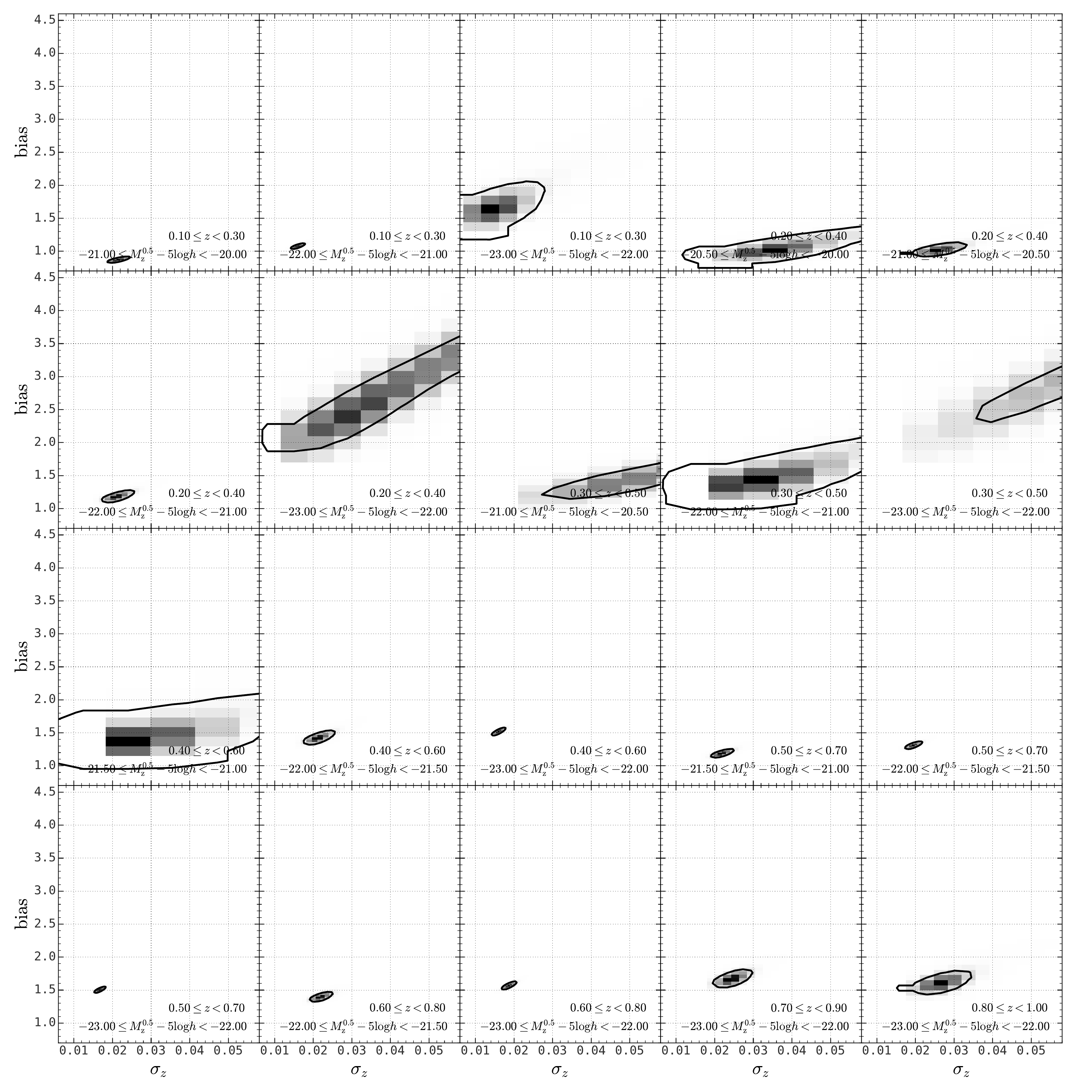}
    \caption{ Similar to Fig. \ref{fig:sigma}, but here for 20 blue galaxy sub-samples.}\label{fig:sigma_blue}
\end{figure*}

\subsection{Results for blue sub-samples}

Similar to those clustering measurements and model fittings  carried out for the 40 all and red galaxy (sub-)samples in section \ref{sec:mf}, we also looked into the 20 blue galaxy sub-samples. Similar to the 20 all galaxy samples, we also measure their $w^{\rm obs}_{p}$ with two different
$r_{\pi,\max} = 50/100\mpch$ values, where the results are shown in Fig. \ref{fig:corr_blue} with open squares.  The overall behaviors are roughly similar to those of all galaxy samples. Here we
can see that the gap between $w^{\rm obs}_{p}$ for different $r_{\pi,\max}$ values are larger in these blue sub-samples. In addition, we find in a few sub-samples, the projected 2PCFs on large scales are significantly boosted. 
 
Following the same procedures for each of the 20 all galaxy samples, we run 10 MCMC chains, each with 50,000 steps for each of our blue galaxy sub-samples. After discarding the  burn-in  stage models, we use the remaining 400,000 MCMC models to describe the  posterior distributions of the model parameters. The results  are shown in Fig. \ref{fig:sigma_blue} with contours. Here again, we show in Fig. \ref{fig:corr_blue} using solid lines the best fitting model 
predictions. Unlike the 40 all and red (sub-)samples, here we can see that 
there are some blue sub-samples where our models can not describe their
behaviors well, especially those with boosted clustering strengths on large
scales.  For these sub-samples, our constraints on the photoz error and bias parameters are 
also quite poor.

Note that even in the framework of a more sophisticated HOD model, the parameters in general can change the shape of $w_p$ on small scales and the amplitude on large scales. Since the clustering of galaxies on large scales is modelled through the combination of a constant galaxy bias and the auto correlation of dark matter particles, the very different  shape of $w_p$ on scales $r_p>3\mpch$ shown in Fig. \ref{fig:corr_blue} will not be well modelled in the HOD models.
We thus believe that there should be some systematic photoz errors or photometry errors  in these blue sub-samples, so that their projected 2PCFs on large scales are boosted and can not be well modelled by theory. Because of these, we omit
further analysis for these 20 blue galaxy (sub-)samples in this study.

\end{appendix}

\end{multicols}

\begin{thebibliography}{99}

\bibitem {LCRS} Shectman, S. A., Landy, S. D., Oemler,
A., et al. ApJ, 470, 172 (1996). 
\bibitem {2dFGRS} Colless, M., Dalton, G., Maddox, S., et al.
 MNRAS, 328, 1039 (2001). 
\bibitem {SDSS} York, D. G., Adelman, J., Anderson, Jr.,
J. E., et al. AJ, 120, 1579 (2000). 
\bibitem {deep2_clustering1} Coil, A. L., Davis, M., Madgwick, D. S.,
et al. ApJ, 609, 525 (2004).
\bibitem {deep2_clustering2} Coil, A. L., Newman, J. A., Croton, D.,
et al. ApJ, 672, 153 (2008).
\bibitem {vipers_clustering} Guzzo, L., Scodeggio, M., Garilli, B.,
et al. A\&A, 566, A108 (2014).
\bibitem {2010dken.book..246B} Bassett, B., \& Hlozek, R. Baryon
acoustic oscillations, 246 (2010).
\bibitem {2012MNRAS.427.2132P} Padmanabhan, N., Xu, X., Eisenstein,
D. J., et al. MNRAS, 427, 2132 (2012).
\bibitem {2014MNRAS.441...24A} Anderson, L., Aubourg, É., Bailey, S., et al. MNRAS, 441, 24 (2014).
\bibitem {2015PhRvD..92l3516A} Aubourg, É., Bailey, S., Bautista, J. E., et al. PhRvD, 92, 123516 (2015).
\bibitem {2019MNRAS.484.3818S} Sarpa, E., Schimd, C., Branchini, E., \&
Matarrese, S. MNRAS, 484, 3818 (2019).
\bibitem {2001PhR...340..291B} Bartelmann, M., \& Schneider, P.
PhR, 340, 291 (2001).
\bibitem {2008ARNPS..58...99H} Hoekstra, H., \& Jain, B. Annual
Review of Nuclear and Particle Science,
58, 99 (2008).
\bibitem {2014MNRAS.442.2017V} Vegetti, S., Koopmans, L. V. E., Auger,
M. W., Treu, T., \& Bolton, A. S. 
MNRAS, 442, 2017 (2014).
\bibitem {2015MNRAS.446.1356H} Han, J., Eke, V. R., Frenk, C. S., et al. MNRAS, 446, 1356 (2015).
\bibitem {2016MNRAS.456.2301W} Wang, W., White, S. D. M.,
Mandelbaum, R., et al. MNRAS,
456, 2301 (2016).
\bibitem {2015JCAP...01..024Z} Zhang, J., Luo, W., \& Foucaud, S. 
JCAP, 2015, 024 (2015).
\bibitem {2017ApJ...836...38L} Luo, W., Yang, X., Zhang, J., et al. 
ApJ, 836, 38 (2017).
\bibitem {2018ApJ...862....4L} Luo, W., Yang, X., Lu, T., et al.
ApJ, 862, 4 (2018).
\bibitem {2019ApJ...874....7D} Dong, F., Zhang, J., Yu, Y., et al. 
ApJ, 874, 7 (2019).
\bibitem {1994ApJ...426...23F} Feldman, H. A., Kaiser, N., \& Peacock,
J. A. ApJ, 426, 23 (1994).
\bibitem {2002MNRAS.330..506H} Hamilton, A. J. S., \& Tegmark, M. 
MNRAS, 330, 506 (2002).
\bibitem {2002MNRAS.335..887T} Tegmark, M., Hamilton, A. J. S., \& Xu,
Y. MNRAS, 335, 887 (2002).
\bibitem {2004ApJ...606..702T} Tegmark, M., Blanton, M. R., Strauss,
M. A., et al. ApJ, 606, 702 (2004).
\bibitem {2011MNRAS.416.3017B} Beutler, F., Blake, C., Colless, M., et al. MNRAS, 416, 3017 (2011).
\bibitem {2016ApJ...833..287L} Li, Z., Jing, Y. P., Zhang, P., \& Cheng,
D. ApJ, 833, 287 (2016).
\bibitem {2001MNRAS.327.1297P} Percival, W. J., Baugh, C. M.,
Bland-Hawthorn, J., et al. MNRAS, 327, 1297 (2001).
\bibitem {2007MNRAS.381.1053P} Percival, W. J., Cole, S., Eisenstein, D. J., et al. MNRAS, 381, 1053 (2007).
\bibitem {2010MNRAS.401.2148P} Percival, W. J., Reid, B. A., Eisenstein, D. J., et al. MNRAS, 401, 2148 (2010).
\bibitem {2005MNRAS.361..824G} Gaztañaga, E., \& Scoccimarro, R.
MNRAS, 361, 824 (2005).
\bibitem {2009ApJ...698..479G} Guo, H.,  Jing, Y. P. ApJ, 698,
479 (2005). 
\bibitem {2009ApJ...702..425G} Guo, H.,  Jing, Y. P. ApJ, 702, 425 (2009).
\bibitem {jing98} Jing, Y. P., Mo, H. J.,  Börner, G. 
ApJ, 494, 1 (1998).
\bibitem {2004MNRAS.350.1153Y} Yang, X., Mo, H. J., Jing, Y. P., van den Bosch, F. C.,  Chu, Y. MNRAS, 350, 1153 (2004).
\bibitem {2005ApJ...633..560E} Eisenstein, D. J., Zehavi, I., Hogg, D. W., et al. ApJ, 633, 560 (2005).
\bibitem {2005ApJ...630....1Z} Zehavi, I., et al. ApJ, 630, 1 (2005).
\bibitem {licheng06} Li, C., Kauffmann, G., Jing, Y. P., et al.
 MNRAS, 368, 21 (2006).
\bibitem {zehavi11} Zehavi, I., Zheng, Z., Weinberg, D. H.,
et al. ApJ, 736, 59 (2011).
\bibitem {guo14} Guo, H., Zheng, Z., Zehavi, I., et al.
 MNRAS, 441, 2398 (2014).
\bibitem {guo15} Guo, H., Zheng, Z., Zehavi, I., et al. MNRAS, 453, 4368 (2015).
\bibitem {2016ApJ...833..241S} Shi, F., Yang, X., Wang, H., et al. 
ApJ, 833, 241 (2016).
\bibitem {2018ApJ...858...30G} Guo, H., Yang, X.,  Lu, Y. ApJ,
858, 30 (2018). 
\bibitem{Xu18} Xu, H., Zheng, Z., Guo, H., et al. MNRAS, 481, 5470 (2018)
\bibitem {2002ApJ...575..587B} Berlind, A. A.,  Weinberg, D. H. 
ApJ, 575, 587 (2002).
\bibitem {zheng05} Zheng, Z., Berlind, A. A., Weinberg,
D. H., et al. ApJ, 633, 791 (2005).
\bibitem {guo16} Guo, H., Zheng, Z., Behroozi, P. S., et al.
 MNRAS, 459, 3040 (2016).
\bibitem {2018MNRAS.478.2019Y} Yuan, S., Eisenstein, D. J.,  Garrison,
L. H. MNRAS, 478, 2019 (2018).
\bibitem {2015MNRAS.454.1161Z} Zu, Y.,  Mandelbaum, R. MNRAS, 454, 1161 (2015).
\bibitem {2016MNRAS.457.4360Z} Zu, Y.,  Mandelbaum, R. MNRAS, 457, 4360 (2016).
\bibitem {2018MNRAS.476.1637Z} Zu, Y.,  Mandelbaum, R. MNRAS, 476, 1637 (2018).
\bibitem {yang03} Yang, X., Mo, H. J.,  van den Bosch,
F. C. MNRAS, 339, 1057 (2003).
\bibitem {2006MNRAS.365..842C} Cooray, A. MNRAS, 365, 842 (2006).
\bibitem {2007MNRAS.376..841V} van den Bosch, F. C., Yang, X., Mo,
H. J., et al. MNRAS, 376, 841 (2007).
\bibitem {yang12} Yang, X., Mo, H. J., van den Bosch, F. C.,
Zhang, Y.,  Han, J. ApJ, 752, 41 (2012).
\bibitem {2015ApJ...799..130R} Rodríguez-Puebla, A., Avila-Reese, V.,
Yang, X., et al. ApJ, 799, 130 (2015).
\bibitem {2006ApJ...647..201C} Conroy, C., Wechsler, R. H.,  Kravtsov,
A. V. ApJ, 647, 201 (2006).
\bibitem {2016MNRAS.460.3647X} Xu, H., Zheng, Z., Guo, H., Zhu, J., 
Zehavi, I. MNRAS, 460, 3647 (2016)
\bibitem {2013ApJ...767..122G} Guo, H., Zehavi, I., Zheng, Z., et al.
ApJ, 767, 122 (2013)
\bibitem {2007ApJ...667..760Z} Zheng, Z., Coil, A. L.,  Zehavi, I. ApJ, 667, 760 (2007)
\bibitem {2009MNRAS.394..929C} Cacciato, M., van den Bosch, F. C., More,
S., et al. MNRAS, 394, 929 (2009)
\bibitem {2013MNRAS.430..767C} Cacciato, M., van den Bosch, F. C., More,
S., Mo, H.,  Yang, X. MNRAS, 430, 767 (2013)
\bibitem {2007ApJS..172...70L} Lilly, S. J., Le Fèvre, O., Renzini, A.,
et al. ApJS, 172, 70 (2007).
\bibitem {Masjedi06} Masjedi, M., Hogg, D. W., Cool, R. J.,
et al. ApJ, 644, 54 (2006).
\bibitem{2009MNRAS.399.2279M} Myers, A. D., White, M.,  Ball, N. M. MNRAS, 399, 2279 (2009)
\bibitem{2011ApJ...731..117H} Hickox, R. C., Myers, A. D., Brodwin, M.,
et al. ApJ, 731, 117 (2011)
\bibitem {wangwenting11} Wang, W., Jing, Y. P., Li, C., Okumura,
T.,  Han, J. ApJ, 734, 88 (2011).
\bibitem{2012A&A...542A...5C} Coupon, J., Kilbinger, M., McCracken,
H. J., et al. A\&A, 542, A5 (2012)
\bibitem {HSC_clustering16} Harikane, Y., Ouchi, M., Ono, Y., et al. ApJ, 821, 123 (2016).
\bibitem{2018PASJ...70S..11H} Harikane, Y., Ouchi, M., Ono, Y., et al. 
Publications of the Astronomical
Society of Japan, 70, S11 (2018)
\bibitem{2018PASJ...70S..33H} He, W., Akiyama, M., Bosch, J., et al. 
Publications of the Astronomical
Society of Japan, 70, S33 (2018)
\bibitem {2019ApJ...879...71W} Wang, Z., Xu, H., Yang, X., et al. 2019,
ApJ, 879, 71 (2019).
\bibitem {Eisenstein03} Eisenstein, D. J. ApJ, 586, 718 (2003).

\bibitem {Planck2018} Planck Collaboration, Aghanim, N.,
Akrami, Y., et al. A\&A, 641, A6 (2020)
\bibitem {2020arXiv201214998Y} Yang, X., Xu, H., He, M., et al.
 arXiv e-prints, arXiv:2012.14998 (2020)
\bibitem {DESI} DESI Collaboration, Aghamousa, A.,
Aguilar, J., et al. ArXiv e-prints, arXiv:1611.00036 (2016)
\bibitem{DESI_image} Dey, A., Schlegel, D. J., Lang, D., et al. 
ArXiv e-prints, arXiv:1804.08657 (2018).
\bibitem {2016ascl.soft04008L} Lang, D., Hogg, D. W.,  Mykytyn, D. The Tractor: Probabilistic
astronomical source detection and
measurement ascl:1604.008 (2016)
\bibitem {2016AAS...22831701B} Blum, R. D., Burleigh, K., Dey, A., et al.
American Astronomical Society
Meeting Abstracts 228, 317.01 (2016).
\bibitem {2017PASP..129f4101Z} Zou, H., Zhou, X., Fan, X., et al. 
PASP, 129, 064101 (2017).
\bibitem {2016AAS...22831702S} Silva, D. R., Blum, R. D., Allen, L., et al.
American Astronomical Society
Meeting Abstracts 228, 317.02 (2016).
\bibitem {2020arXiv200106018Z} Zhou, R., Newman, J. A., Mao, Y.-Y.,
et al. arXiv e-prints,
arXiv:2001.06018 (2020).
\bibitem {2002ApJ...571..172Z} Zehavi, I., Blanton, M. R., Frieman, J. A.,
et al. ApJ, 571, 172 (2002)
\bibitem {2005ApJ...629..143B}
Blanton, M. R., Eisenstein, D., Hogg,
D. W., Schlegel, D. J.,  Brinkmann, J.
ApJ, 629, 143 (2005)
\bibitem {Breiman2001} Breiman, L. Machine Learning, 45,
5 (2001).
\bibitem {2012ApJS..199...26H} Huchra, J. P., Macri, L. M., Masters, K. L., et al. ApJS, 199, 26 (2012)
\bibitem {2009MNRAS.399..683J} Jones, D. H., Read, M. A., Saunders, W.,
et al. MNRAS, 399, 683 (2009)
\bibitem {2007AJ....133..734B} Blanton, M. R.,  Roweis, S. AJ,
133, 734 (2007)
\bibitem {correlation_function} Landy, S. D.,  Szalay, A. S. ApJ,
412, 64 (1993). 
\bibitem {1998ApJ...494L..41S} Szapudi, I.,  Szalay, A. S. ApJ,
494, L41 (1998).
\bibitem {2000ApJ...535L..13K} Kerscher, M., Szapudi, I.,  Szalay, A. S.
 ApJL, 535, L13 (2000).
\bibitem {10.1007/978-981-13-7729-7_1} Sinha, M.,  Garrison, L. 
Software Challenges to Exascale
Computing, ed. A. Majumdar 
R. Arora (Singapore: Springer
Singapore), 3â€0. (2019).
\bibitem {2020MNRAS.491.3022S} Sinha, M.,  Garrison, L. H. 
MNRAS, 491, 3022 (2020).
\bibitem {2017ApJ...846...61G} Guo, H., Li, C., Zheng, Z., et al.
ApJ, 846, 61 (2017)
\bibitem {CAMB} Lewis, A., Challinor, A.,  Lasenby, A.
 ApJ, 538, 473 (2000).
\bibitem {2009MNRAS.396...19N} Norberg, P., Baugh, C. M., Gaztañaga,
E.,  Croton, D. J.  MNRAS, 396,
19 (2009).
\bibitem {emcee} Foreman-Mackey, D., Hogg, D. W., Lang,
D.,  Goodman, J. PASP, 125, 306 (2013).
\bibitem {2016MNRAS.460.1371B} Beck, R., Dobos, L., Budavári, T., Szalay,
A. S.,  Csabai, I. MNRAS, 460,1371 (2016).
\bibitem {2019ApJS..242....8Z} Zou, H., Gao, J., Zhou, X.,  Kong, X.
 ApJS, 242, 8 (2019).
\bibitem {2013A&A...557A..17M} Marulli, F., Bolzonella, M., Branchini, E.,
et al. A\&A, 557, A17 (2013).
\bibitem {2018ApJ...853...69C} Cowley, W. I., Caputi, K. I., Deshmukh,
S., et al. ApJ, 853, 69 (2018).
\bibitem {VIPERS_data1} Moutard, T., Arnouts, S., Ilbert, O., et al.
 A\&A, 590, A102 (2016).
\bibitem {2016MNRAS.458.4015Z} Zheng, Z.,  Guo, H. MNRAS, 458,
4015 (2016).








\end{thebibliography}
\end{document}